\documentclass[aps,prb,twocolumn,superscriptaddress]{revtex4-1}

\usepackage{graphicx}
\usepackage{amsmath}
\usepackage{amssymb}
\usepackage{wasysym}
\usepackage{textcomp}
\usepackage{color}

\definecolor{lila}{rgb}{0.5,0,1}
\newcommand{\bnen}{\begin{equation}}
\newcommand{\eden}{\end{equation}}
\newcommand{\bean}{\begin{eqnarray}}
\newcommand{\eean}{\end{eqnarray}}

\newcommand{\bna}{\begin{array}}
\newcommand{\eda}{\end{array}}

\begin{document}

\title{Interaction quench and thermalization in a one-dimensional topological Kondo insulator}

\author{I. Hagym\'asi}
\affiliation{Department of Physics,
Arnold Sommerfeld Center for Theoretical Physics (ASC),
Fakult\"{a}t f\"{u}r Physik, Ludwig-Maximilians-Universit\"{a}t M\"{u}nchen,
D-80333 M\"{u}nchen, Germany}
\affiliation{Munich Center for Quantum Science and Technology (MCQST), Schellingstr. 4, D-80799 M\"unchen, Germany}
\affiliation{Strongly Correlated Systems "Lend\"ulet" Research Group, Institute for Solid State
Physics and Optics, MTA Wigner Research Centre for Physics, Budapest H-1525 P.O. Box 49, Hungary
}

\author{C. Hubig}
\affiliation{Max-Planck-Institut f\"ur Quantenoptik,
Hans-Kopfermann-Strasse 1, 85748 Garching, Germany}

\author{U. Schollw\"ock}
\affiliation{Department of Physics,
Arnold Sommerfeld Center for Theoretical Physics (ASC),
Fakult\"{a}t f\"{u}r Physik, Ludwig-Maximilians-Universit\"{a}t M\"{u}nchen,
D-80333 M\"{u}nchen, Germany}
\affiliation{Munich Center for Quantum Science and Technology (MCQST), Schellingstr. 4, D-80799 M\"unchen, Germany}

\date{\today}

\begin{abstract}
We study the nonequilibrium dynamics of a one-dimensional topological Kondo insulator, modelled by 
a $p$-wave Anderson lattice model, following a quantum quench of the on-site interaction strength. 
Our goal is to examine how the quench influences the 
topological properties of the system, therefore our main focus is the 
time evolution of the string order parameter, entanglement spectrum and the topologically-protected 
edge states.
We point out that postquench local observables can be well captured by a 
thermal ensemble up to a certain 
interaction strength. Our 
results demonstrate 
that the topological properties after the interaction quench are preserved. 
Though the absolute value of the string order parameter decays 
in time, the analysis of the entanglement spectrum, Loschmidt echo and the 
edge states indicates the 
robustness of the topological properties in the time-evolved state.
These predictions 
could be directly tested in state-of-the-art cold-atom experiments.
\end{abstract}

\maketitle
\section{Introduction} 
The time evolution in closed many-body quantum systems has attracted enormous 
attention due to their unusual thermalization properties.\cite{Rigol2008,RevModPhys.83.863,
Eisert2015}  For a large class of quantum
systems the eigenstate thermalization hypothesis\cite{Rigol2008,PhysRevE.50.888,PhysRevA.43.2046} 
provides a way to understand the thermalization of local observables. On the other hand, the 
topological phases typically cannot be characterized by a local order parameter but by a nonlocal 
one.\cite{RevModPhys.89.041004} A paradigmatic example of a symmetry-protected topological phase is 
the Haldane phase of spin-1 Heisenberg model on a chain, where a hidden diluted antiferromagnetic 
order can be described by a nonlocal string order 
parameter.\cite{HALDANE1983464,PhysRevLett.50.1153,PhysRevB.40.4709}  While the 
time evolution of local observables has been investigated 
extensively over the last years, much less is known about the time-dependent properties of string 
operators. In recent works\cite{PhysRevB.94.024302,PhysRevB.90.020301,PhysRevB.96.214206} this 
question has been addressed for both spin and bosonic models. It has also been 
shown very recently that the topological phase may abruptly disappear during 
the unitary time evolution even if certain symmetry protecting the phase is 
present in the quench Hamiltonian.\cite{PhysRevLett.121.090401}

These findings motivate our present work, we examine what happens, 
when a topological phase is realized with 
fermions to account for the charge fluctuations missing in a purely spin-based model.  To this end 
we consider an Anderson lattice model with $s$- and $p$-wave electrons with a nonlocal hybridization
 term.\cite{PhysRevB.92.205128} 
This model originates from the $p$-wave Kondo-Heisenberg model\cite{Coleman2014}
suggested by Alexandrov and Coleman  to capture the topology and strong correlations simultaneously 
behind the alleged topological Kondo insulating material, SmB$_6$.\cite{Fiskprb2013,Paglioneprx2013,
Kim2013} The latter model has attracted significant attention: Abelian bosonization revealed that 
its ground state is actually a Haldane phase,\cite{Galitskiprx2015} later on this finding triggered 
further research and with the help of several other techniques including the density matrix 
renormalization-group (DMRG)\cite{PhysRevB.93.165104,PhysRevB.92.205128,PIP336,PhysRevB.96.075124,
pillay2018topological} and quantum Monte Carlo methods,\cite{Zhong2017} the existence of a Haldane-
like ground state was confirmed, going beyond the limits of bosonization. The $p$-wave Anderson and 
Kondo lattices are related to each other via a Schrieffer-Wolff transformation, by which one can 
eliminate the charge degrees of freedom of the $s$ electrons in the Anderson lattice model.\cite{
PhysRevB.92.205128} 
 The $p$-wave Anderson lattice may be experimentally realized  by loading ultracold fermions into 
$p$-band optical lattices.\cite{PhysRevB.92.205128,PhysRevB.96.075124}

While significant work has been done to explore the ground-state properties, including the effect 
of perturbations\cite{PhysRevB.93.165104,pillay2018topological} and even finite temperature effects,
\cite{zhong2017finite} much less is known about the nonequilibrium properties of 1D topological 
Kondo insulators. Our goal in this paper is to fill this gap by studying the time-dependent 
properties of the Haldane phase emerging in the $p$-wave Anderson lattice model, when an interaction
 quench is applied which is well-controlled experimentally using Feshbach resonances.\cite{
RevModPhys.80.885} We study the relaxation and thermalization of various quantities, namely, the 
double occupancy, spin correlations and we also consider the string order parameter, entanglement 
spectrum, Loschmidt echo and the edge states for revealing the properties of 
the time-evolved topological state. 
The unitary time evolution is performed using the matrix-product-state based 
time-dependent variational principle (TDVP) 
method.\cite{PhysRevLett.107.070601,PhysRevB.94.165116}  
Nevertheless, the maximal 
time reachable in our simulation is limited by the entanglement growth,\cite{SCHOLLWOCK201196} and 
in global quenches like the present one, the entanglement grows linearly in time.\cite{Fazio2006} 

The paper is organized as follows. In Sec.~II our model is introduced together 
with the applied 
methods. In Sec.~III A our results are presented for local observables of the 
model following the 
interaction quench, then in Sec.~III B nonlocal quantities (string order, 
entanglement spectrum, Loschmidt echo) 
characterizing the topological order are studied together with the edge states in the nonequilibrium
 case. Finally, in Sec.~IV we give the conclusions of this work.

\section{Model and methods}
\begin{figure}[!h]
\includegraphics[scale=0.32]{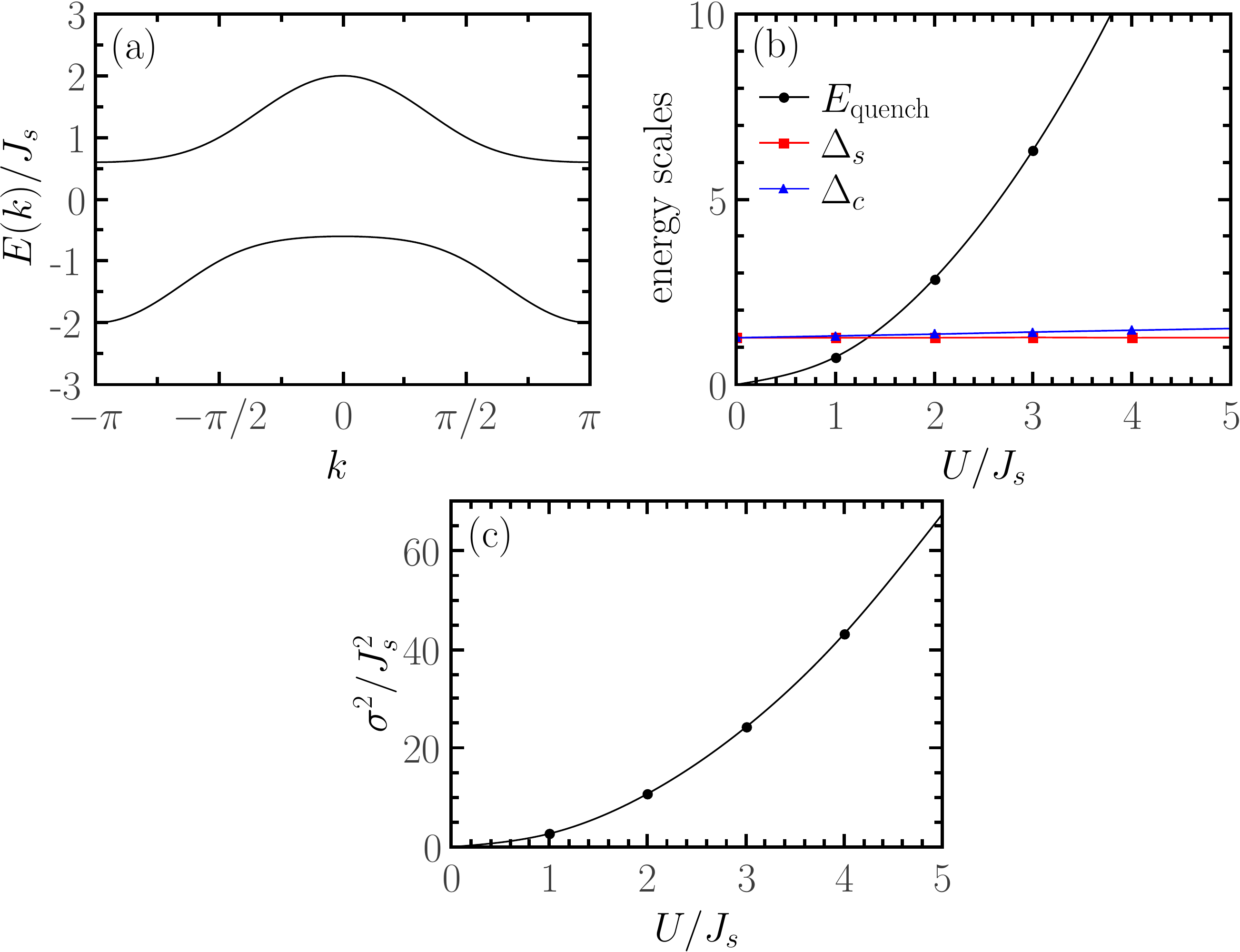}
\caption{(a) The noninteracting ($\hat{\mathcal{H}}_U\equiv 0$) band structure, 
$E(k)$, of the Hamiltonian defined in Eq.~(\ref{eq:Hamiltonian}). (b) The energy 
of the quench and the charge and spin gaps (in units of 
$J_s$) as a function of the Hubbard interaction for $L=80$. (c) The variance  
(Eq.~(\ref{eq:variance})) of the initial state with respect to the quench 
Hamiltonian  for $L=80$.}
\label{fig:energy-scales}
\end{figure}

The $p$-wave Anderson Hamiltonian can be written as follows:
\begin{equation}
\label{eq:Hamiltonian}
\hat{\mathcal{H}}=\hat{\mathcal{H}}_s+\hat{\mathcal{H}}_p+\hat{\mathcal{H}}_{sp}+\hat{\mathcal{H}}_
U,
\end{equation}
where  $\hat{\mathcal{H}}_s$ and $\hat{\mathcal{H}}_p$ describe two tight-binding chains with $s$- 
and $p$-wave symmetries, respectively:
\begin{equation}
\begin{split}
\hat{\mathcal{H}}_s=J_s\sum_{j=1}^{L-1}\sum_{\sigma}(\hat{s}_{j\sigma}^{\dagger}
\hat{s}_{j+1\sigma}^{\phantom\dagger}+ {\rm H.c.}),\\
\hat{\mathcal{H}}_p=-J_p\sum_{j=1}^{L-1}\sum_{\sigma}(\hat{p}_{j\sigma}^{\dagger
}\hat{p}_{j+1\sigma}^{
\phantom\dagger}+ {\rm H.c.}),
\end{split}
\end{equation}
where $J_s$ and $J_p$ are the hopping amplitudes of the corresponding orbitals, since we use $t$ 
for denoting time.
The different symmetries of the two subsystems are encoded in the hybridization 
term, that is, only a nonlocal hybridization can be present 
which is described by the term $\hat{\mathcal{H}}_{sp}$:
\begin{equation}
\hat{\mathcal{H}}_{sp}=J_{sp}\sum_{j=1}^{L}\sum_{\sigma}\left[\hat{s}_{j\sigma}^{\dagger}(\hat{p}_{
j+1\sigma}^{\phantom\dagger}-\hat{p}_{j-1\sigma}^{\phantom\dagger}) + {\rm H.c.}\right],
\end{equation}
where $J_{sp}$ is the hybridization matrix element and $\hat{p}_{j\sigma}^{\phantom\dagger}$ ($\hat{
s}_{j\sigma}^{\phantom\dagger}$) annihilates a fermion with $p$- ($s$)-wave 
symmetry. Furthermore 
$\hat{p}_{0\sigma}^{\phantom\dagger}=\hat{p}_{L+1\sigma}^{\phantom\dagger}=0$ is 
assumed. 
Finally
\begin{equation}
\hat{\mathcal{H}}_U=E_s\sum_{j=1}^{L}\sum_{\sigma} \hat{n}^s_{j\sigma}+
U\sum_{j=1}^{L}\hat{n}^s_{j\uparrow}\hat{n}^s_{
j\downarrow}
\end{equation}
contains the on-site energy, $E_s$, and the Hubbard interaction, $U$, 
associated to the $s$-wave chain. We consider the half-filled case, that is, 
there are two electrons per site, altogether $N=2L$ electrons in the system. 
The on-site energy of the $s$-wave chain is set to $E_s=-U/2$ (symmetric case), 
which guarantees that the local occupancy of both orbitals is one.
We set $J_s$ 
as the energy unit, $\hbar=k_B=1$, furthermore $J_{sp}/J_s=1$ and 
$J_p/J_s=\pi/10$. Our choice of the hopping parameters is motivated by the fact 
that in the $U\rightarrow+\infty$ limit, where the Kondo lattice case is 
recovered, the velocities of the gapless excitations in the Heisenberg and the 
tight-binding chains coincide, hence the effect of the hybridization (which 
introduces the nontrivial topology in the system) is more 
emphasized.\cite{PhysRevB.92.205128}
The hopping amplitudes are assumed to have the same sign ($J_sJ_p>0$), 
which ensures that the noninteracting ground state is always a band insulator, 
the band structure is shown in Fig.~\ref{fig:energy-scales}(a) for our choice 
of the parameters.
In addition, it can also be classified as a $Z_2$ band insulator due to the 
special form of the hybridization term.\cite{Zhong2017} If the hopping 
amplitudes had opposite signs, the ground state would be metallic and the 
topological reasoning would not make sense.
The ground-state properties of the model have been studied recently, and it 
turned out that the noninteracting 
ground state is adiabatically connected to the interacting one,\cite{PhysRevB.96.075124} that is, 
no topological phase transition takes place as $U$ is switched on. In the present work we address 
the scenario that the system is prepared in the initially noninteracting ground state:
\begin{equation}
\hat{\mathcal{H}}(U_i=0)|\Psi_0\rangle=E_0|\Psi_0\rangle
\end{equation}
assuming that the ground state has no net magnetic moments, and then we evolve it with the 
interacting Hamiltonian:
\begin{equation}
|\Psi(t)\rangle=e^{-i\hat{\mathcal{H}}(U_f=U)t} |\Psi_0\rangle.
\end{equation}
In what follows $\langle\dots\rangle(t)$ denotes expectation value over $|\Psi(t)\rangle$.
\par The time evolution is performed using the TDVP 
method,\cite{PhysRevLett.107.070601,PhysRevB.94.165116} which does not 
require a manual partition of the Hamiltonian into non-overlapping parts and we can avoid the 
Trotter-Suzuki decomposition of the time-evolution operator and the use of swap 
gates. On the other hand it introduces a projection error but this is 
much smaller than the truncation error (which is controlled during 
the simulation), since the time evolution is started from a fairly entangled 
state. In our 
simulations the total discarded weight was set to $10^{-7}$, and the largest bond dimension used 
was $\sim6000$. We considered chains with lengths $L=40-80$ and show results for 
system size $L=80$ (unless stated otherwise) 
for which the finite-size effects were negligible. We compared runs with different total discarded 
weights and show only data that are indistinguishable on the scale of the figures. The ground-state 
calculations were performed using the standard DMRG procedure,\cite{White:DMRG1,White:DMRG2,
schollwock2005,hallberg2006,hubig15:_stric_dmrg} while finite-temperature 
calculations were obtained with the ancilla 
method.\cite{PhysRevLett.93.207204}

\section{Results}
Before diving into the details of the quench dynamics, it is instructive to 
look at how the low-energy charge and spin excitations relate to the energy of 
the quench. Since we consider chains with open boundary conditions, we must 
adopt a different definition of the spin and charge gap to rule out the gapless 
edge modes in the system:
\begin{equation}
 \begin{split}
   \Delta_s(L)&=E_0(2,2L)-E_0(0,2L),\\
 \Delta_c(L)&=E_0(0,2L+4)-E_0(0,2L),
 \end{split}
\end{equation}
where $E_0(T^z,N)$ is the ground-state energy with total magnetization $T^z$ 
and number of electrons, $N$. 
The definition for the spin gap is analogous to the definition of the Haldane 
gap in spin systems. Similar considerations apply for the charge gap, namely,
at half filling the edge modes already host two fermions and can host up to 
four fermions altogether, thus, we need to add four fermions to the system to 
obtain a bulk excitation, while keeping the total magnetization zero. The 
energy of the quench by definition is:
\begin{equation} 
E_{\mathrm{quench}}(U)=\langle\Psi_{0}|\hat{\mathcal{H}}
(U)|\Psi_0\rangle-\langle\Psi_U|\hat { \mathcal{H}}(U)|\Psi_U\rangle,
\end{equation}
where $|\Psi_U\rangle$ denotes the ground state of $\hat{\mathcal{H}}(U)$.
These quantities are plotted together in Fig.~\ref{fig:energy-scales}(b).
For weak quenches, $U \lesssim 2$, the quench does not really probe the higher 
lying excitations; however, above this value the energy of the quench becomes 
much larger than the first excitations in the spin and charge sectors that are 
roughly constant as $U$ is increased. Besides the quench energy, it is also 
instructive to calculate the variance, $\sigma^2$ of the initial state with 
respect to the quench Hamiltonian:
\begin{equation}
\label{eq:variance}
 \sigma^2(U)=\langle\Psi_{0}|\hat{\mathcal{H}}^2
(U)|\Psi_0\rangle-\langle\Psi_0|\hat { \mathcal{H}}(U)|\Psi_0\rangle^2.
\end{equation}
This enables us to estimate what fraction of excitations 
takes part in the quench. The variance is shown in 
Fig.~\ref{fig:energy-scales}(c) and increases as $\sigma^2\propto U^2$. Based 
on 
these observation, we may expect 
qualitatively different behavior for $U \lesssim 2$ and $U \gtrsim 2$.
\subsection{Local observables}
\begin{figure}[!h]
\includegraphics[scale=0.8]{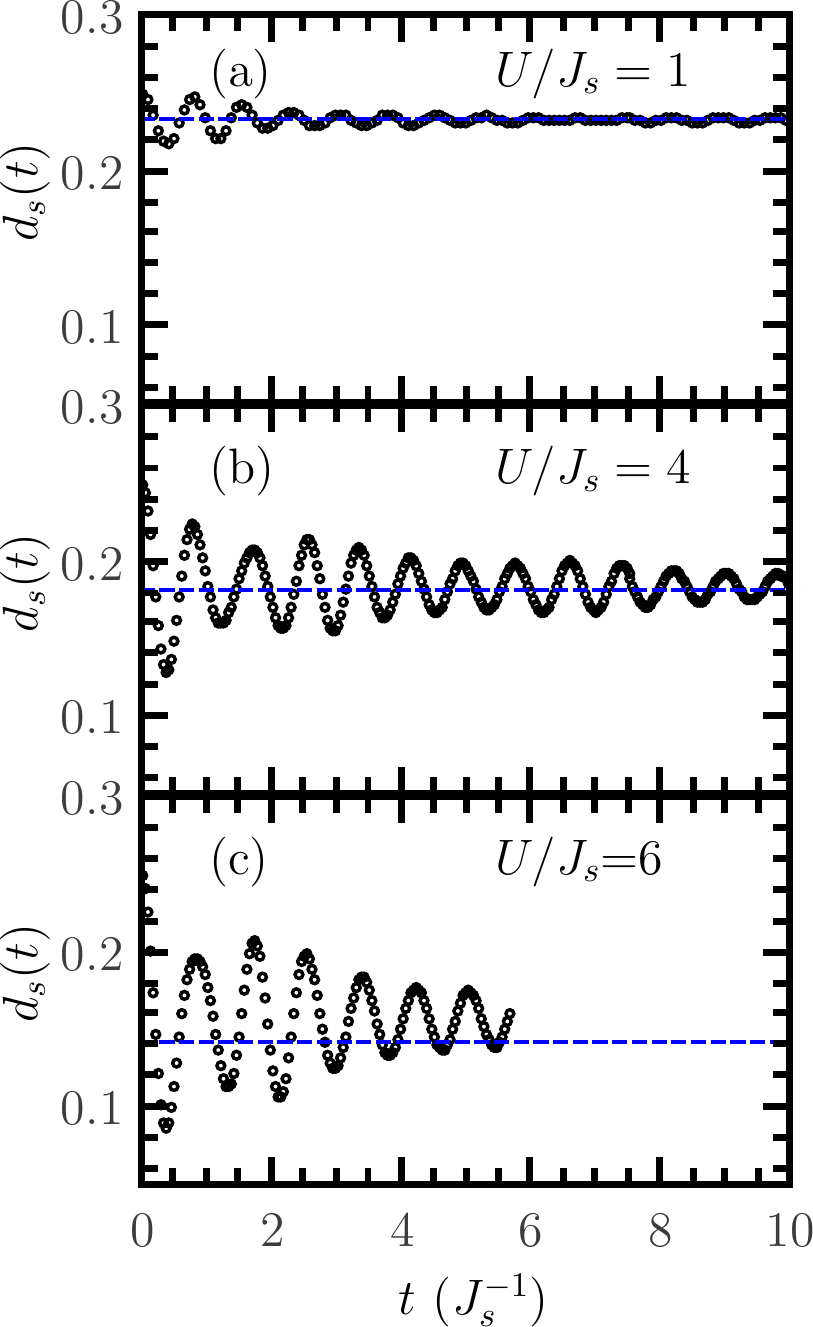}
\caption{Time evolution of double occupancy on the $s$-wave chain after the interaction quench from 
$U_i=0$ to $U_f=U$ as indicated in the figures. In case of $U/J_s=6$ we were not able to go beyond 
$t\approx 6 J_s^{-1}$ due to entanglement growth. The dashed lines denote the 
corresponding thermal 
averages. }
\label{fig:double_occupancy}
\end{figure}

\begin{figure}[!h]
\includegraphics[scale=0.6]{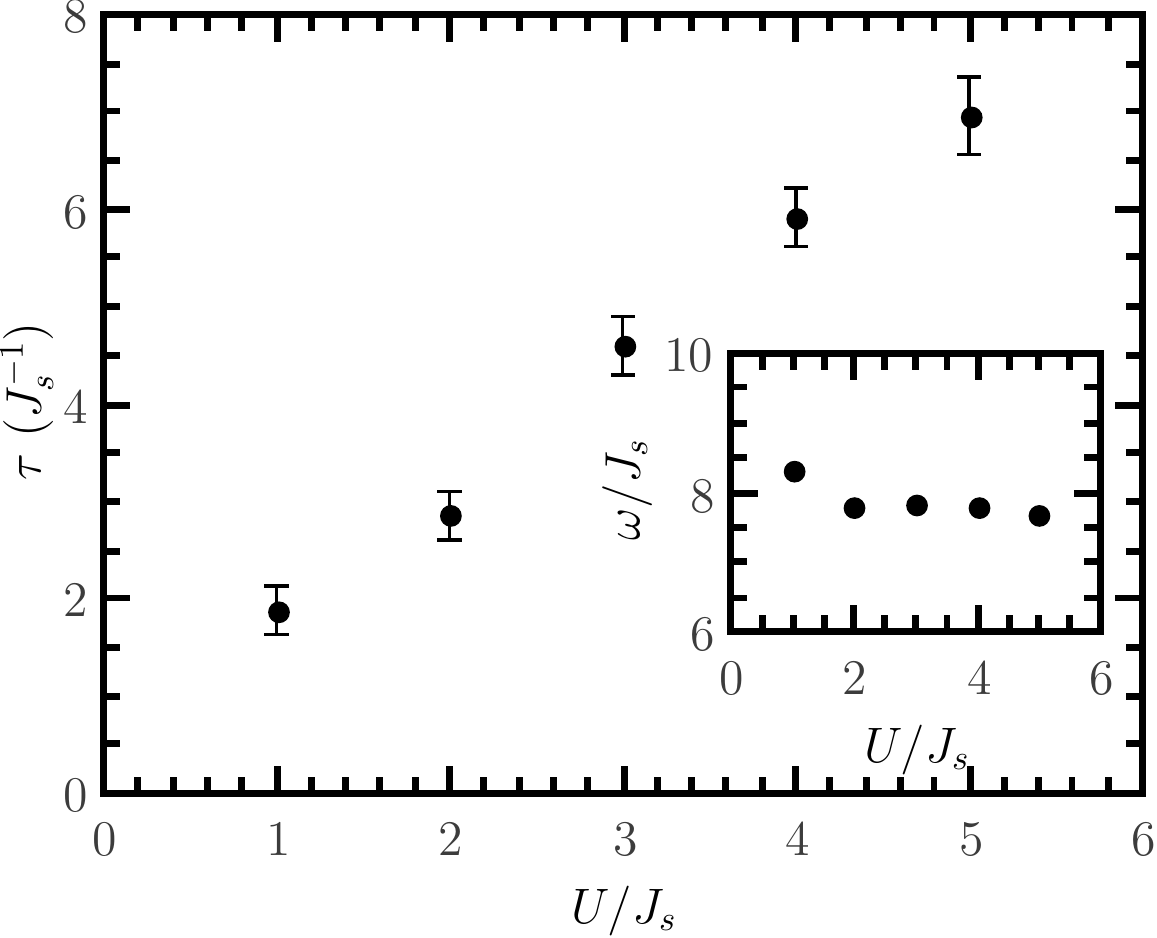}
\caption{The main and inset figures show the relaxation time, $\tau$, and  
the frequency of the oscillation, $\omega$, respectively, as a function of the 
Hubbard interaction. The error
bars are the asymptotic standard error resulting from the least-squares fit
of Eq.~(\ref{eq:fit-function}) to the data.}
\label{fig:fitting-parameters}
\end{figure}

First, we investigate the time evolution of the double occupancy on the $s$-wave chain:
\begin{equation}
d^s(t)=\frac{1}{L}\sum_{j=1}^L\left\langle \hat{n}^s_{j\uparrow} 
\hat{n}^s_{j\downarrow}\right\rangle(t),
\end{equation}
since this quantity is readily accessible in quantum gas experiments.\cite{PhysRevLett.110.205301,
PhysRevLett.104.080401}
The time evolution of $d^s$ is shown in Figs.~\ref{fig:double_occupancy} (a)-(c) following the 
interaction quenches from $U_i=0$ to $U_f=U$.
Since the system is initially prepared in an uncorrelated state, the double occupancy at $t=0$ is 
very close to $1/4$ although the hybridization between the two orbitals is present.  We observe 
that the data can be fitted reasonably well with the function
\begin{equation}
d^s(t)=a\sin(\omega t+\phi)\exp(-t/\tau)+\bar{d}^s.
\label{eq:fit-function}
\end{equation}
For strong quenches ($U/J_s\gtrsim 3$) we discarded the transient behavior for $t\lesssim 2$ in the 
fitting procedure. To characterize the postquench dynamics it is worth 
investigating how the fitting parameters depend on the model parameters. We 
could reach long enough times up to $U/J_s=5$ to reliably use the fitting 
function. The results are shown in Fig.~\ref{fig:fitting-parameters}.
We observe that the relaxation time increases linearly with the Hubbard 
interaction strength, which is perfectly consistent with the \emph{a priori} 
expectations concluded from Fig.~\ref{fig:energy-scales}, since for large 
interaction strength the quench drives the system far away from the equilibrium 
ground state and the slower the system relaxes the larger the Hubbard 
interaction strength is. On the other, the frequency of the oscillation 
do not exhibit any significant dependence on the interaction strength, it 
remains roughly constant, $\omega\approx 8J_s$.

\begin{figure}[h]
\includegraphics[scale=0.6]{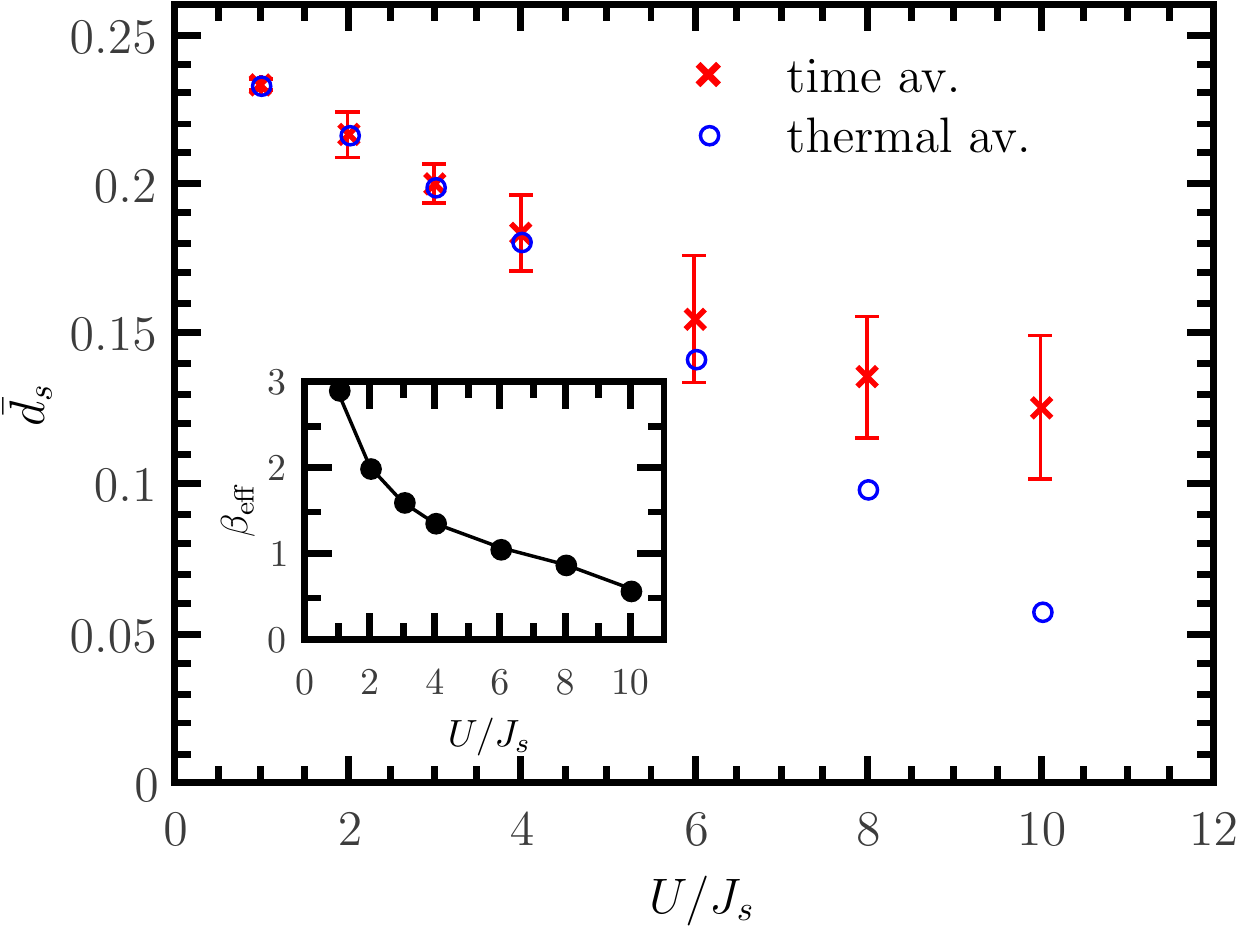}
\caption{Time and thermal average of the double occupancy on the $s$-wave chain as a function of 
the postquench interaction value. The error bars show the standard deviation from the mean value. 
The inset figure shows the effective inverse temperature as a function of the postquench interaction
 value.}
\label{fig:nb-gs-time-av}
\end{figure}

One can also extract the time average of the double occupancy, $\bar{d}^s$ from 
the fit results or by 
averaging the above data for $t\gtrsim 2/J_s$. The latter one is used for 
calculating the time-averaged quantities later on. To address the question of 
thermalization, we compare them with the corresponding thermal averages in 
Fig.~\ref{fig:nb-gs-time-av}.
The thermal ensemble is defined by the density matrix 
$\hat{\rho}(\beta)=e^{-\beta_{\rm eff} \hat{\mathcal{H}}}/\mathcal{Z}$, where 
$\mathcal{Z}$ is the partition function and the effective inverse 
temperature, $\beta_{\rm eff}$, is determined from the following relation:
\begin{equation}
\label{eq:invtemp}
\langle\Psi_0| \hat{\mathcal{H}}(U_f=U) |\Psi_0\rangle = {\rm Tr} 
\left[\hat{\mathcal{H}}(U_f=U) \hat{\rho}(\beta) \right].
\end{equation}
It is readily seen that the postquench time averages are in a very good agreement with the thermal 
averages  corresponding to the postquench $U$ as long as $U$ is relatively weak. These results 
suggest that the double occupancy thermalizes for $U/J_s\lesssim 6$, however, for $U/J_s\gtrsim 6$ 
a discrepancy is observed indicating a nonthermal value. A possible explanation 
can be that the thermalization time is much longer than the time reachable in 
our simulation, and the time averages in our time window are different from 
those in the steady state. The inverse effective temperature 
satisfying Eq.~(\ref{eq:invtemp})  as a function of the postquench $U$ is shown 
in the inset of Fig.~\ref{fig:nb-gs-time-av}, where the expected divergence for 
$U\rightarrow0$ is visible.
\par
\begin{figure}[!t]
\includegraphics[scale=0.8]{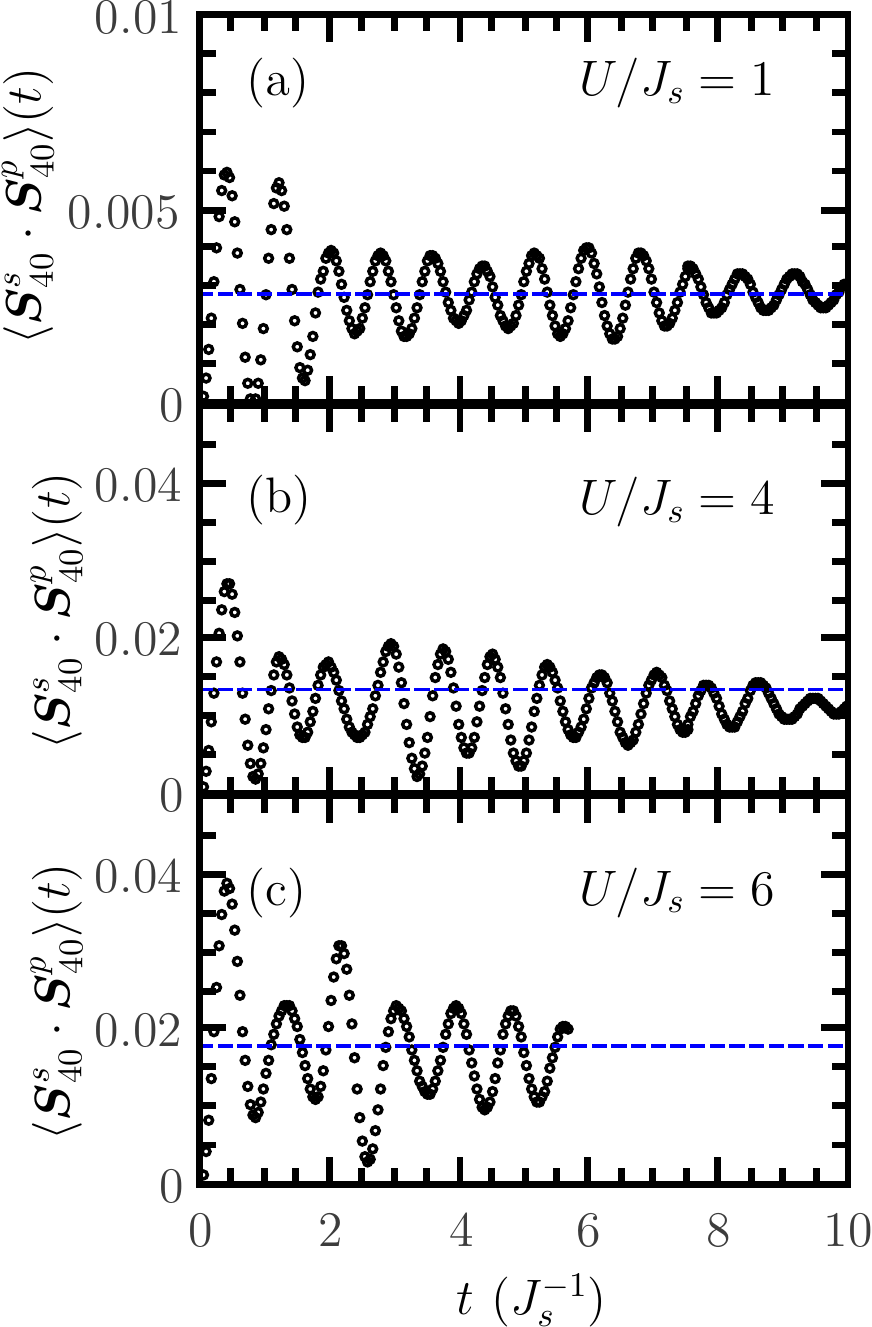}
\caption{Time evolution of local spin correlation between the $s$ and $p$ electrons measured in the 
middle of the $L=80$ chain, after the interaction quench from $U_i=0$ to $U_f=U$ as indicated in 
the figures. The dashed lines denote the 
corresponding thermal 
averages.}
\label{fig:spin-corr}
\end{figure}

 It is also instructive to study the local spin correlations between the $s$ and $p$ electrons, 
which is shown in Fig.~\ref{fig:spin-corr} together with the corresponding 
thermal averages.
The spin operators for fermion species $a\in\{s,p\}$ are defined as
\begin{equation}
\hat{\boldsymbol{S}}_j^a=\frac{1}{2}\sum_{\beta\gamma}\hat{a}^{\dagger}_{j\beta}\boldsymbol{\sigma}
^{\phantom\dagger}_{\beta\gamma}\hat{a}^{\phantom\dagger}_{j\gamma},
\end{equation}
where $\boldsymbol{\sigma}$ is the vector of Pauli matrices.
Initially the correlation between the two types of electrons is zero due to the uncorrelated state, 
then ferromagnetic correlation develops similarly to the equilibrium case in the presence of 
interaction. The emergence of local ferromagnetic correlations can be 
understood from the following argument. Switching on the interaction results in 
antiferromagnetic nearest-neighbor correlations among the $s$ electrons, that 
is, $\langle \hat{\boldsymbol{S}}_j^s \hat{\boldsymbol{S}}_{j+1}^s\rangle < 0$. 
The correlation between nearest-neighbor $s$ and $p$ electrons are also 
antiferromagnetic, $\langle \hat{\boldsymbol{S}}_j^s 
\hat{\boldsymbol{S}}_{j+1}^p\rangle < 0$, since the hybridization term, which 
connects these sites, favors the formation of a singlet. (In the conventional 
Anderson lattice, this hybridization is on-site and prefers to have a 
local Kondo singlet.) We can repeat the same argument for sites $(j-1,j)$, from 
which one can quickly see that the correlation $\langle \hat{\boldsymbol{S}}_j^s 
\hat{\boldsymbol{S}}_{j}^p\rangle$ should be ferromagnetic. Thus the two 
$S=1/2$ fermions in the lattice form a $S=1$ object in each site, which are 
coupled antiferromagnetically. This is also the reason why the present system 
resembles to the Haldane phase.
Regarding the thermalization, it also exhibits similarities to the 
double occupancy; 
for $U/J_s\lesssim 3$ the time-averages agree well with those of the thermal 
ensemble. The discrepancy at larger values of $U$ can be explained by the 
previous argument for the thermalization time.

\subsection{Nonlocal observables and edge states}
\begin{figure}[!h]
\includegraphics[scale=0.6]{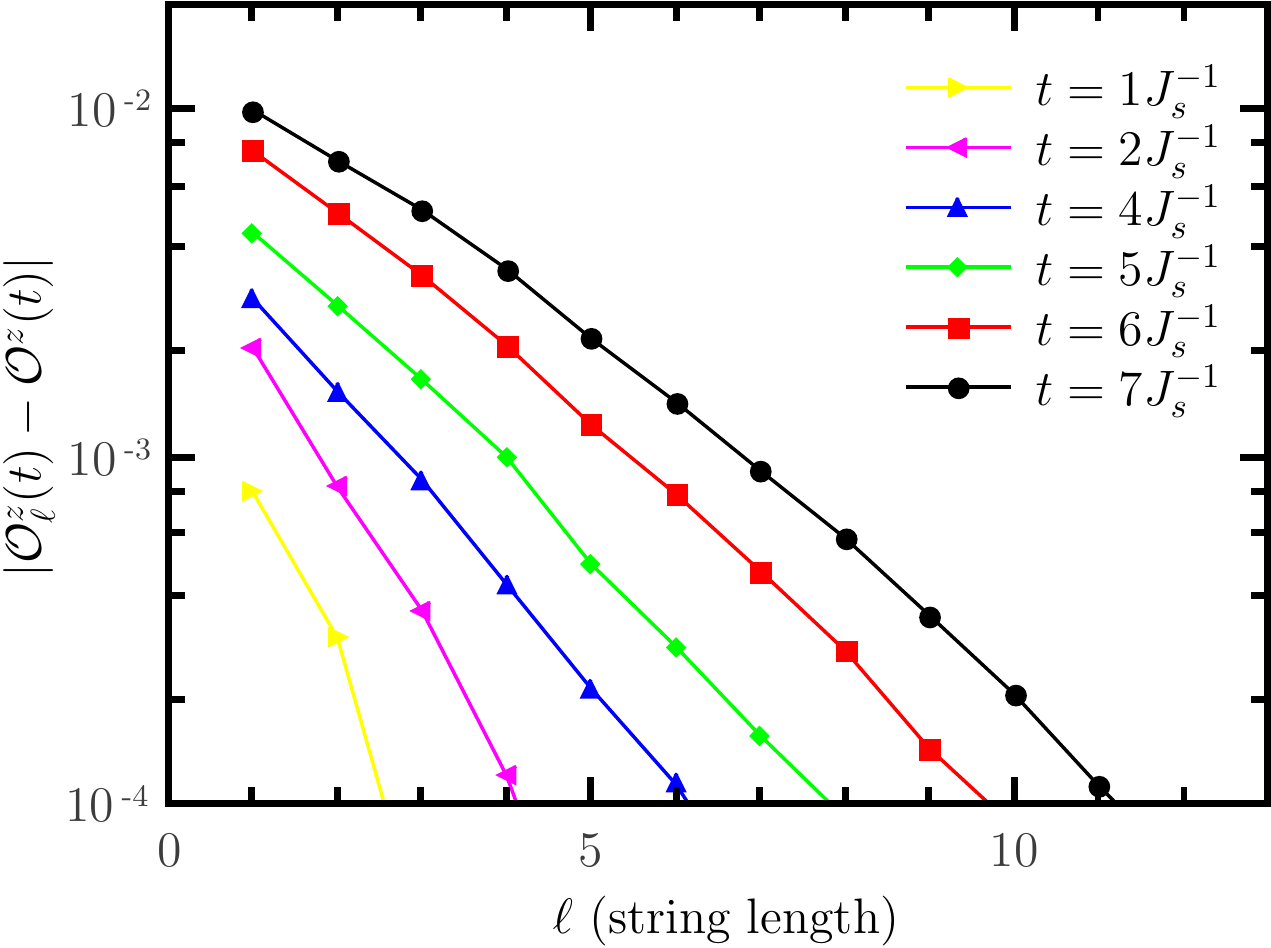}
\caption{Deviation of string correlations from their thermodynamic value as a 
function of string 
length measured at different times on  a log-lin scale for $U_f/J_s=5$. }
\label{fig:subtracted-string}
\end{figure}

\begin{figure}[h]
\includegraphics[scale=0.8]{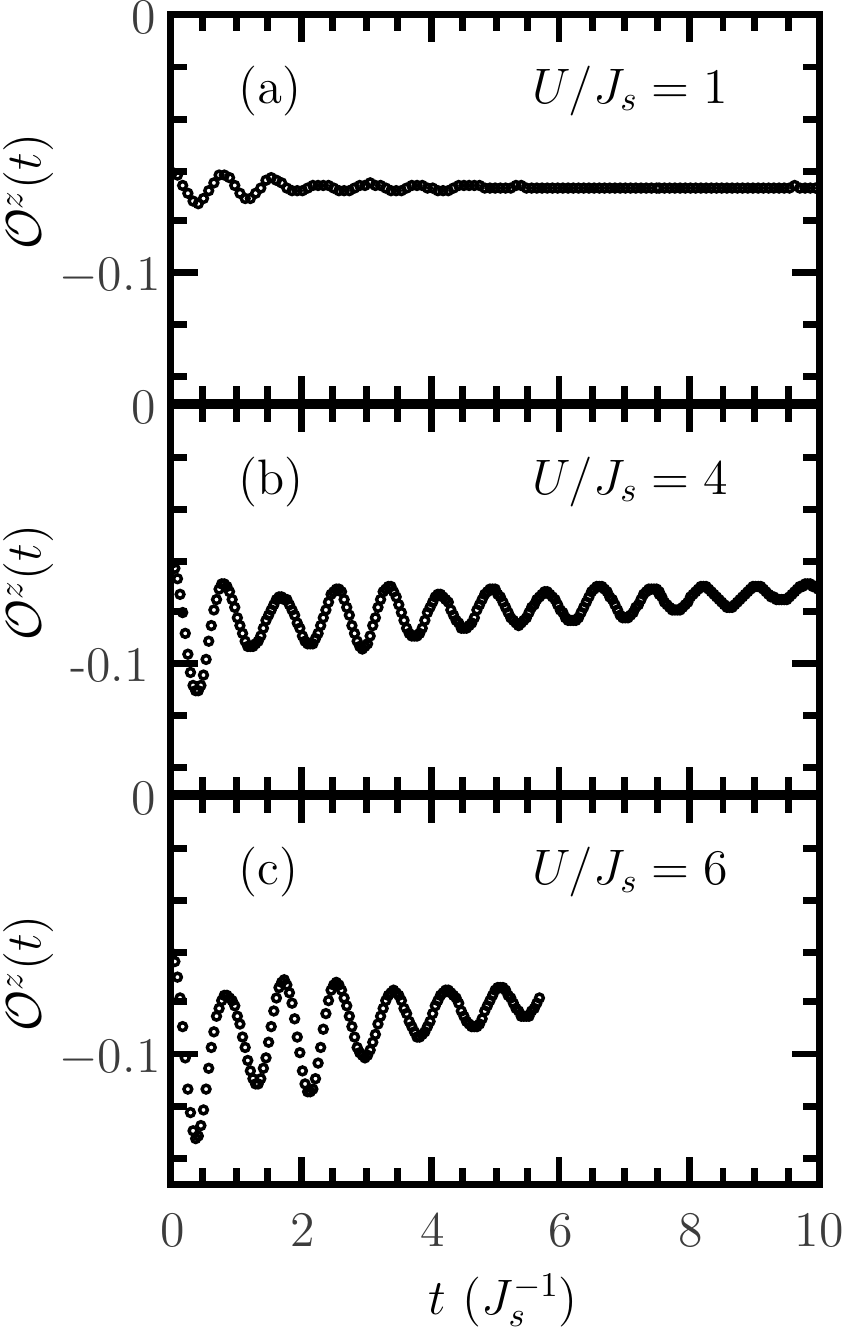}
\caption{Time evolution of string order parameter following the interaction quench from $U_i=0$ to 
$U_f=U$ as indicated in the figures.}
\label{fig:string-op}
\end{figure}

In what follows we focus on the behaviour of nonlocal quantities following the 
interaction quench. 
Previously it was shown, that the noninteracting ground state is adiabatically connected to the 
interacting case\cite{PhysRevB.96.075124} (both being in the Haldane phase); however, it is not 
trivial what happens to its topological properties when the interaction is abruptly turned on. The 
Haldane phase is generically characterized by the breaking of a 
hidden $Z_2\times Z_2$ symmetry, which implies a symmetry-protected 
topological order manifesting itself in (i) an evenly degenerate entanglement spectrum, (ii) 3 
nonvanishing string order parameters and (iii) a ground-state degeneracy depending on the boundary 
conditions.\cite{Pollmann2010prb,Pollmann2011prb,Pollmann2012prb} In this subsection we address the 
time-dependent properties of the entanglement spectrum, string order parameter and the edge states. 
The entanglement spectrum, $\Lambda_j$, is immediately accessed by performing a 
Schmidt decomposition of the wave function into two halves:
\begin{equation}
\left|\Psi\right\rangle=\sum_j \Lambda_j |\phi_j\rangle_A |\phi^{\prime}_j\rangle_B.
\end{equation}
The presence of the diluted antiferromagnetic order is characterized by the string operator:
\begin{equation}
\label{eq:string-op}
\hat{\mathcal{O}}^{\alpha}_{\ell}= \hat{T}_j^{\alpha} \left[ \prod_{n=j+1}^{j+\ell-1}e^{i\pi \hat{T}
_n^{\alpha}} \right] \hat{T}_{j+\ell}^{\alpha}  \quad (\alpha\in \{x,y,z\}).
\end{equation}
The ground state of a system exhibits string order when the string order 
parameter, $\mathcal{O}^{\alpha}$ fulfills
\begin{equation}
\mathcal{O}^{\alpha}=\lim_{\ell\rightarrow\infty}\left\langle 
\hat{\mathcal{O}}^{\alpha}_{\ell} \right\rangle \neq 0
\end{equation}
for any $\alpha$, or alternatively in the time-dependent case:
\begin{equation}
\mathcal{O}^{\alpha}(t)=\lim_{\ell\rightarrow\infty}\left\langle 
\hat{\mathcal{O}}^{\alpha}_{\ell} \right\rangle(t) \neq 0.
\end{equation}
In Eq.~(\ref{eq:string-op}) $\hat{T}_j^{\alpha}$ is the appropriate component of the total spin 
operator at site $j$:
\begin{equation}
\hat{\boldsymbol{T}}_j=\hat{\boldsymbol{S}}_j^s+\hat{\boldsymbol{S}}_j^p.
\end{equation}
Due to the $\mathrm{SU}(2)$ symmetry of the Hamiltonian (\ref{eq:Hamiltonian}), 
it is sufficient to consider 
one of the three 
string order operators, therefore we concentrate on 
$\hat{\mathcal{O}}^{z}_{\ell}$ in the following.
\begin{figure}[!h]
\includegraphics[scale=0.6]{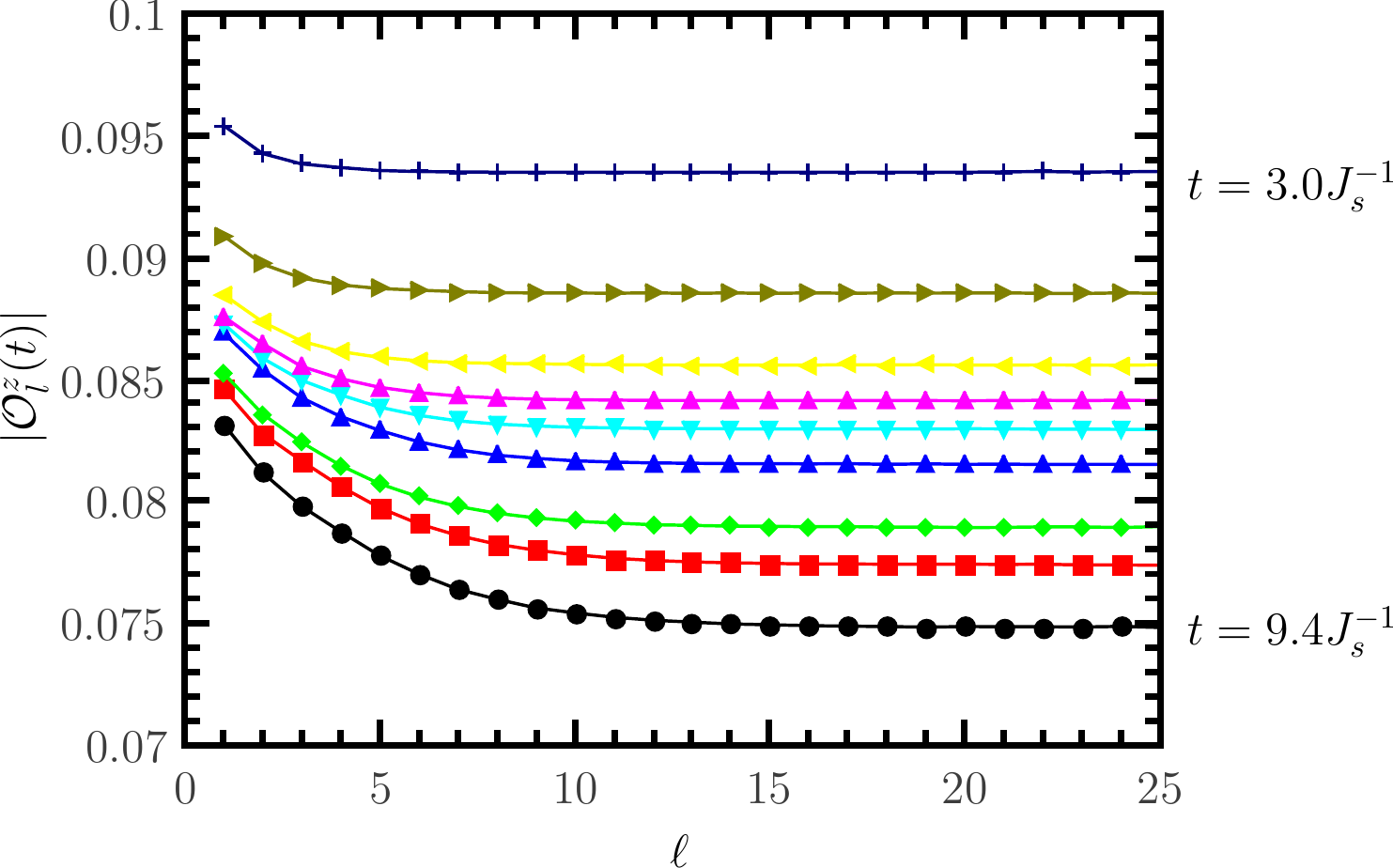}
\caption{String correlations after the interaction quench from $U_i=0$ to 
$U_f/J_s=4$ as a function of length $\ell$. Each line corresponds to a 
different time ranging from $t=3J_s^{-1}$ (top) to $t=9.4J_s^{-1}$ (bottom) 
with time spacing $0.8J_s^{-1}$.}
\label{fig:string-length-U4}
\end{figure}
We first discuss the behavior of the string operator for various lengths and 
times ($t>3J_s^{-1}$ to exclude the transient behavior at short times), which 
is shown in Fig.~\ref{fig:string-length-U4}. It is observed that the string 
correlations start decreasing after the quench. It is also immediately seen 
that $\mathcal{O}^z(t)$ is approached exponentially as the string length is 
increased, furthermore, the more time has elapsed the slower the expectation 
value of the string 
operator reaches its thermodynamic value, which is demonstrated by 
Fig.~\ref{fig:subtracted-string}.
Next we turn our attention to the string order 
parameter, shown in Fig.~\ref{fig:string-op} after different 
interaction quenches.
In agreement with the previous finding,\cite{PhysRevB.96.075124} the system exhibits string order 
even for $U=0$. The string order parameter remains nonzero after the quench as 
well but its absolute value starts decreasing, which might vanish 
in the steady state at $t\rightarrow+\infty$, but longer times are out of 
reach due to entanglement growth. This feature is more emphasized for 
stronger quenches, that is, $U/J_s\gtrsim4$.
For weak interaction quenches this behavior is not observed, which may 
originate from the fact that the defect density is low, thus, the 
thermalization time may be very long and the decay is not visible at this time 
scale.
\par
\begin{figure}[!h]
\includegraphics[scale=0.8]{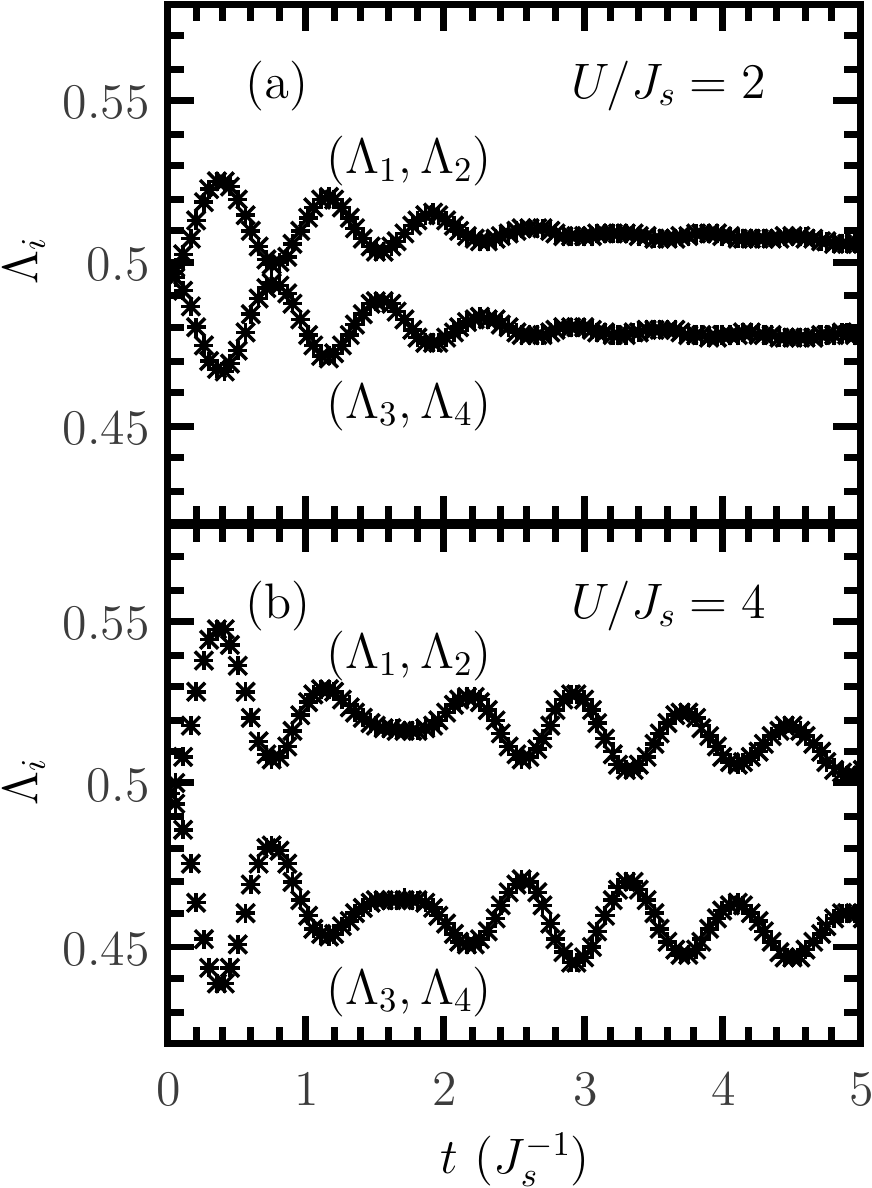}
\caption{Time evolution of the low-lying entanglement spectrum after the interaction quench from $U_
i=0$ to $U_f=U$. The consecutive Schmidt values are denoted by $\times$ and 
$+$, respectively.}
\label{fig:entanglement-spect}
\end{figure}

 It is important to note; however, that the string order parameter is a 
basis-dependent quantity, and its decrease or alleged disappearance is not 
sufficient evidence for the destruction of the topological properties. 
Therefore it is also intriguing to analyze the entanglement spectrum after the 
quench, which is another hallmark of symmetry-protected topological phases and 
basis-independent. 
For better visibility we 
consider only the largest 4 Schmidt values, which are plotted in Fig.~\ref{fig:entanglement-spect}, 
but the higher lying values also exhibit qualitatively similar behavior.
It is immediately observed that the initially fourfold degenerate Schmidt value becomes twofold 
degenerate 
following the interaction quench. As we would expect from the nonzero 
string order parameter, 
the 
degeneracy of the spectrum is also preserved for finite times. The crossover  
of the fourfold degeneracy into two twofold degenerate branches after the 
quench is analogous to what happens when one consider 
the 
evolution of the entanglement spectrum of the ground state as a function of $U$. 
\par
\begin{figure}[!t]
\includegraphics[scale=0.6]{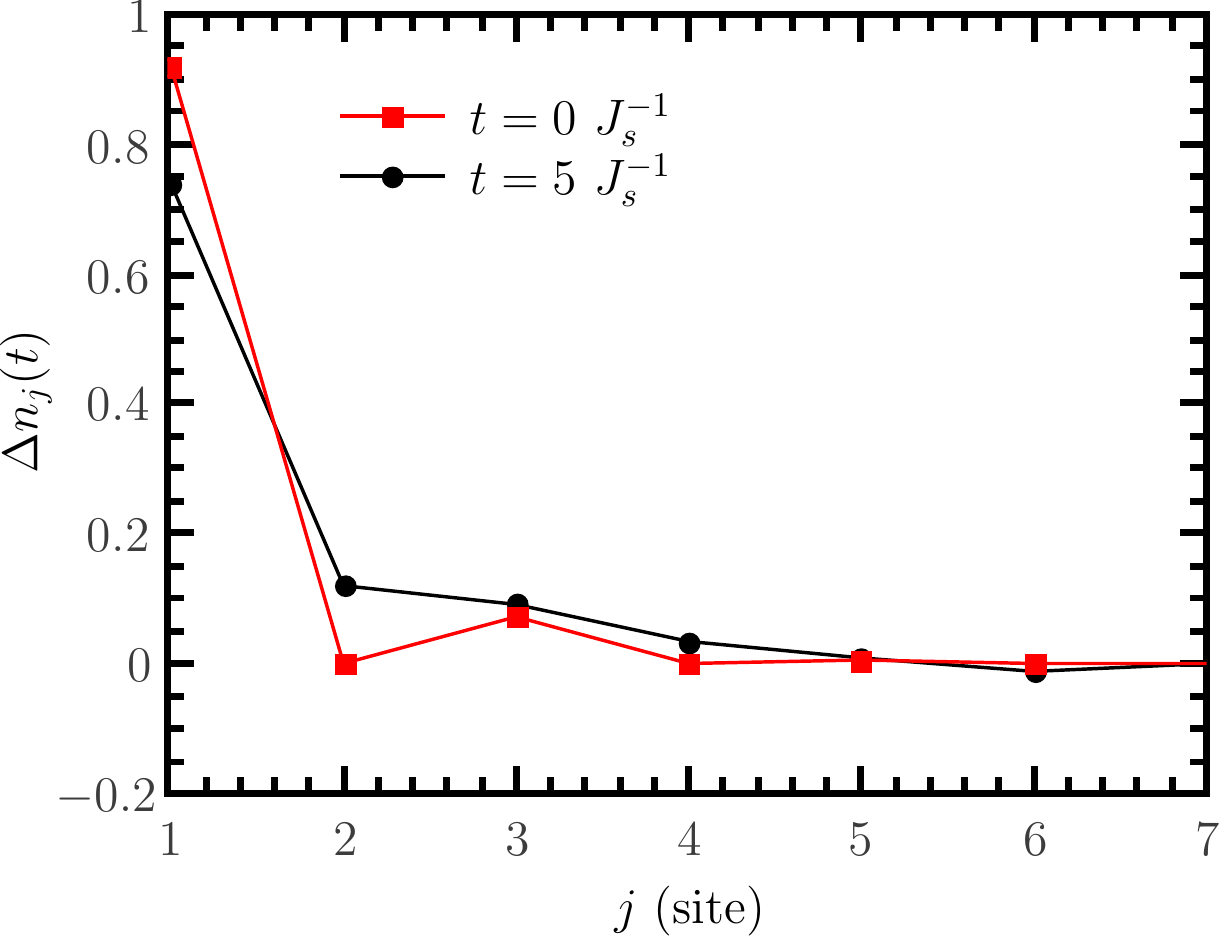}
\caption{Charge profile at the left edge of the system in the initial state and after a quench with 
$U_f/J_s=4$.}
\label{fig:charge-profil}
\end{figure}

 Based on the splitting in the entanglement 
spectrum at $U=0$, one may think that the edge states of the steady state 
should exhibit similar 
behavior as the ground 
state does for finite $U$. Namely, the ground state for $U=0$ and $\sum_j T_j^z=0$ has a holon and 
a doublon 
edge state resulting in a vanishing spin profile, but a nonuniform charge 
profile at the edges.\cite{PhysRevB.96.075124} For $U>0$ these 
edge states become excited states, while the spin-1/2 edge states possess lower energy hence it 
results in a 
uniform charge distribution and an accumulation of 1/2 spins at the edges, forming a singlet. To 
see what happens in the quenched states, we investigated the difference in the spatial charge 
profile, $\Delta n_j$ (Fig.~\ref{fig:charge-profil}), defined as:
\begin{equation}
\Delta n_j(t)= \left\langle n_j \right\rangle(t) -n_0,
\end{equation}
where $n_j=n_j^s+n_j^p$ is the total particle number operator at site $j$ and $n_0=2$ is the 
average occupancy per site in the half-filled case.
Surprisingly, the charge edge states appear to be frozen during the interaction quench and the spin 
profile remains identically zero (not shown) despite the fact that there is 
a finite $U$ present in 
the system. This fact  clearly indicates that the quenched system will preserve the topological 
order at finite times but its properties are different from what one would 
naively expect. 
\par
\begin{figure}[!t]
\includegraphics[scale=0.6]{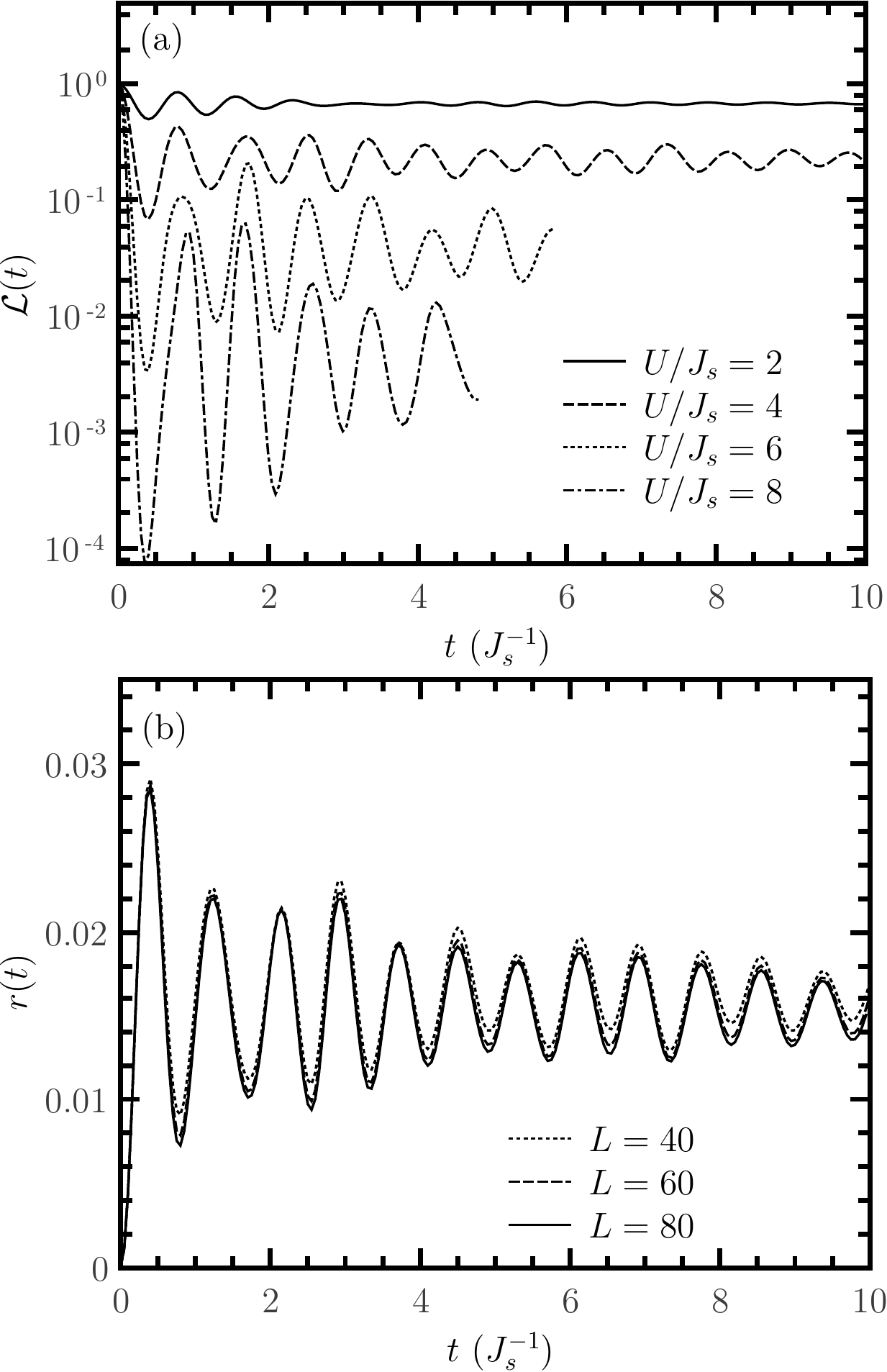}
\caption{(a) Loschmidt echo (using a log-lin scale) after the quench 
with different values of $U$, and $L=40$. (b) The rate function for $U/J_s=4$ 
and different system sizes.}
\label{fig:loschmidt-echo}
\end{figure}

 Due to the fact that the time-evolved state exhibits similar topological 
properties as the initial state, it is interesting to consider the Loschmidt 
echo during the time evolution:
\begin{equation}
 \mathcal{L}(t)=|\langle\Psi_0 |\Psi(t)\rangle|^2,
\end{equation}
which precisely quantifies the deviation of the time-evolved state from initial 
one.
This is shown in Fig.~\ref{fig:loschmidt-echo}(a) for several values of the 
Hubbard interaction strength.
It is observed that for weak interaction quenches ($U/J_s\lesssim 2$), the 
Loschmidt echo is fairly large $\mathcal{L}(t)\sim 0.7$. This may not surprise 
us if we recall Fig.~\ref{fig:energy-scales}(b), that is, the quench energy is 
comparable with the energy of the low-lying excitations, meaning that the 
system remains close the initial state. What is more remarkable is that the 
Loschmidt echo saturates to a value of $\mathcal{L}(t)\sim 0.2$ even for 
$U/J_s=4$, when the quench pushes the system far away from the ground state, 
and similarly, it also oscillates around a finite value for other Hubbard 
interaction strengths. Since the Loschmidt echo, in general, is expected to 
decay exponentially in time in ergodic systems, we conclude that the quench 
does not drive the system to completely explore the Hilbert space, but it 
remains trapped in a region close to the initial state, in spite of the fact 
that the quench energy is quite large compared to the gaps in the system. One 
can naturally ask if the above statements based on the Loschmidt echo hold in 
the thermodynamic limit. Since the Loschmidt echo itself is not applicable for 
infinite system size, one usually introduces the rate function, $r(t)$:
\begin{equation}
 r(t)=-\frac{1}{L}\log[\mathcal{L}(t)],
\end{equation}
which has a well-defined thermodynamic limit. We calculated this quantity for 
different chain lengths in Fig.~\ref{fig:loschmidt-echo}(b) to address the 
finite-size effects. We can observe that $r(t)$ exhibits a weak size-dependence 
(in agreement with the short correlation length from 
Fig.~\ref{fig:subtracted-string}), which supports our arguments based on 
the Loschmidt echo.
\section{Conclusions}
We have presented a numerical analysis of an interaction quench in a 1D topological Kondo insulator 
modelled by a 
$p$-wave Anderson lattice model with nonlocal hybridization. We studied the time evolution and 
thermalization of different observables: double 
occupancy and local spin correlations. In addition we addressed the behavior of 
several other quantities, including the string order parameter and entanglement 
spectrum directly related to the topological properties.
 In case of double occupancy and local spin correlations we found that the 
thermalization already occurs in our simulation up to interaction 
strength $U/J_s\sim 6$ and $U/J_s\sim 3$, respectively, while for stronger 
quenches the thermalization time is expected to be much longer, which accounts 
for the difference between the time and thermal averages.
 \par Then we turned our attention to the topological properties of the 
system. We pointed out that the topological order is preserved in the 
time-evolved state. Although the decreasing value of the string order parameter 
at first glance would
indicate that the steady state might possesses a trivial 
topology, this can be ruled out by examining the entanglement spectrum and 
Loschmidt echo, which are basis independent quantities unlike the string order 
parameter. 
  We demonstrated that the entanglement spectrum preserves its doubly 
degenerate property and the initial charge edge states remain frozen during the 
time 
evolution instead of the appearance of magnetic edge states. Moreover, the 
Loschmidt echo tends to a finite value during the time evolution, clearly 
indicating that the time-evolved state remains in the same phase.
 \par Our results could be directly tested in cold atom experiments, since the charge profile or 
double occupancies can be routinely measured,\cite{PhysRevLett.110.205301,PhysRevLett.104.080401,
Cheuk1260,Parsons1253,PhysRevLett.116.175301} moreover, the string correlations have also been 
extracted in cutting-edge experiments.\cite{Hilker484} Since the interaction can be varied using 
Feshbach resonances, the presented quench scheme could also be experimentally realized. 
\acknowledgments{We acknowledge fruitful discussions with \"O.~Legeza, 
I.~McCulloch and F.~Pollmann.  I.H.~was supported by the Alexander 
von Humboldt Foundation and in part by Hungarian 
National Research,
Development and Innovation Office (NKFIH) through Grant No. K120569 and the Hungarian Quantum 
Technology National Excellence
Program (Project No.  2017-1.2.1-NKP-2017-00001). C.H.~acknowledges funding 
through ERC Grant QUENOCOBA, ERC-2016-ADG (Grant no. 742102). This work was 
also supported in part by the Deutsche Forschungsgemeinschaft (DFG, German 
Research
Foundation) under Germany’s Excellence Strategy -- EXC-2111 -- 390814868.}

\bibliography{tki_refs.bib}

\begin{thebibliography}{45}%
\makeatletter
\providecommand \@ifxundefined [1]{%
 \@ifx{#1\undefined}
}%
\providecommand \@ifnum [1]{%
 \ifnum #1\expandafter \@firstoftwo
 \else \expandafter \@secondoftwo
 \fi
}%
\providecommand \@ifx [1]{%
 \ifx #1\expandafter \@firstoftwo
 \else \expandafter \@secondoftwo
 \fi
}%
\providecommand \natexlab [1]{#1}%
\providecommand \enquote  [1]{``#1''}%
\providecommand \bibnamefont  [1]{#1}%
\providecommand \bibfnamefont [1]{#1}%
\providecommand \citenamefont [1]{#1}%
\providecommand \href@noop [0]{\@secondoftwo}%
\providecommand \href [0]{\begingroup \@sanitize@url \@href}%
\providecommand \@href[1]{\@@startlink{#1}\@@href}%
\providecommand \@@href[1]{\endgroup#1\@@endlink}%
\providecommand \@sanitize@url [0]{\catcode `\\12\catcode `\$12\catcode
  `\&12\catcode `\#12\catcode `\^12\catcode `\_12\catcode `\%12\relax}%
\providecommand \@@startlink[1]{}%
\providecommand \@@endlink[0]{}%
\providecommand \url  [0]{\begingroup\@sanitize@url \@url }%
\providecommand \@url [1]{\endgroup\@href {#1}{\urlprefix }}%
\providecommand \urlprefix  [0]{URL }%
\providecommand \Eprint [0]{\href }%
\providecommand \doibase [0]{http://dx.doi.org/}%
\providecommand \selectlanguage [0]{\@gobble}%
\providecommand \bibinfo  [0]{\@secondoftwo}%
\providecommand \bibfield  [0]{\@secondoftwo}%
\providecommand \translation [1]{[#1]}%
\providecommand \BibitemOpen [0]{}%
\providecommand \bibitemStop [0]{}%
\providecommand \bibitemNoStop [0]{.\EOS\space}%
\providecommand \EOS [0]{\spacefactor3000\relax}%
\providecommand \BibitemShut  [1]{\csname bibitem#1\endcsname}%
\let\auto@bib@innerbib\@empty
\bibitem [{\citenamefont {Rigol}\ \emph {et~al.}(2008)\citenamefont {Rigol},
  \citenamefont {Dunjko},\ and\ \citenamefont {Olshanii}}]{Rigol2008}%
  \BibitemOpen
  \bibfield  {author} {\bibinfo {author} {\bibfnamefont {M.}~\bibnamefont
  {Rigol}}, \bibinfo {author} {\bibfnamefont {V.}~\bibnamefont {Dunjko}}, \
  and\ \bibinfo {author} {\bibfnamefont {M.}~\bibnamefont {Olshanii}},\ }\href
  {http://dx.doi.org/10.1038/nature06838} {\bibfield  {journal} {\bibinfo
  {journal} {Nature}\ }\textbf {\bibinfo {volume} {452}},\ \bibinfo {pages}
  {854 EP } (\bibinfo {year} {2008})}\BibitemShut {NoStop}%
\bibitem [{\citenamefont {Polkovnikov}\ \emph {et~al.}(2011)\citenamefont
  {Polkovnikov}, \citenamefont {Sengupta}, \citenamefont {Silva},\ and\
  \citenamefont {Vengalattore}}]{RevModPhys.83.863}%
  \BibitemOpen
  \bibfield  {author} {\bibinfo {author} {\bibfnamefont {A.}~\bibnamefont
  {Polkovnikov}}, \bibinfo {author} {\bibfnamefont {K.}~\bibnamefont
  {Sengupta}}, \bibinfo {author} {\bibfnamefont {A.}~\bibnamefont {Silva}}, \
  and\ \bibinfo {author} {\bibfnamefont {M.}~\bibnamefont {Vengalattore}},\
  }\href {\doibase 10.1103/RevModPhys.83.863} {\bibfield  {journal} {\bibinfo
  {journal} {Rev. Mod. Phys.}\ }\textbf {\bibinfo {volume} {83}},\ \bibinfo
  {pages} {863} (\bibinfo {year} {2011})}\BibitemShut {NoStop}%
\bibitem [{\citenamefont {Eisert}\ \emph {et~al.}(2015)\citenamefont {Eisert},
  \citenamefont {Friesdorf},\ and\ \citenamefont {Gogolin}}]{Eisert2015}%
  \BibitemOpen
  \bibfield  {author} {\bibinfo {author} {\bibfnamefont {J.}~\bibnamefont
  {Eisert}}, \bibinfo {author} {\bibfnamefont {M.}~\bibnamefont {Friesdorf}}, \
  and\ \bibinfo {author} {\bibfnamefont {C.}~\bibnamefont {Gogolin}},\ }\href
  {http://dx.doi.org/10.1038/nphys3215} {\bibfield  {journal} {\bibinfo
  {journal} {Nat. Phys.}\ }\textbf {\bibinfo {volume} {11}},\ \bibinfo {pages}
  {124 EP } (\bibinfo {year} {2015})}\BibitemShut {NoStop}%
\bibitem [{\citenamefont {Srednicki}(1994)}]{PhysRevE.50.888}%
  \BibitemOpen
  \bibfield  {author} {\bibinfo {author} {\bibfnamefont {M.}~\bibnamefont
  {Srednicki}},\ }\href {\doibase 10.1103/PhysRevE.50.888} {\bibfield
  {journal} {\bibinfo  {journal} {Phys. Rev. E}\ }\textbf {\bibinfo {volume}
  {50}},\ \bibinfo {pages} {888} (\bibinfo {year} {1994})}\BibitemShut
  {NoStop}%
\bibitem [{\citenamefont {Deutsch}(1991)}]{PhysRevA.43.2046}%
  \BibitemOpen
  \bibfield  {author} {\bibinfo {author} {\bibfnamefont {J.~M.}\ \bibnamefont
  {Deutsch}},\ }\href {\doibase 10.1103/PhysRevA.43.2046} {\bibfield  {journal}
  {\bibinfo  {journal} {Phys. Rev. A}\ }\textbf {\bibinfo {volume} {43}},\
  \bibinfo {pages} {2046} (\bibinfo {year} {1991})}\BibitemShut {NoStop}%
\bibitem [{\citenamefont {Wen}(2017)}]{RevModPhys.89.041004}%
  \BibitemOpen
  \bibfield  {author} {\bibinfo {author} {\bibfnamefont {X.-G.}\ \bibnamefont
  {Wen}},\ }\href {\doibase 10.1103/RevModPhys.89.041004} {\bibfield  {journal}
  {\bibinfo  {journal} {Rev. Mod. Phys.}\ }\textbf {\bibinfo {volume} {89}},\
  \bibinfo {pages} {041004} (\bibinfo {year} {2017})}\BibitemShut {NoStop}%
\bibitem [{\citenamefont {Haldane}(1983{\natexlab{a}})}]{HALDANE1983464}%
  \BibitemOpen
  \bibfield  {author} {\bibinfo {author} {\bibfnamefont {F.}~\bibnamefont
  {Haldane}},\ }\href {\doibase https://doi.org/10.1016/0375-9601(83)90631-X}
  {\bibfield  {journal} {\bibinfo  {journal} {Phys. Lett. A}\ }\textbf
  {\bibinfo {volume} {93}},\ \bibinfo {pages} {464 } (\bibinfo {year}
  {1983}{\natexlab{a}})}\BibitemShut {NoStop}%
\bibitem [{\citenamefont {Haldane}(1983{\natexlab{b}})}]{PhysRevLett.50.1153}%
  \BibitemOpen
  \bibfield  {author} {\bibinfo {author} {\bibfnamefont {F.~D.~M.}\
  \bibnamefont {Haldane}},\ }\href {\doibase 10.1103/PhysRevLett.50.1153}
  {\bibfield  {journal} {\bibinfo  {journal} {Phys. Rev. Lett.}\ }\textbf
  {\bibinfo {volume} {50}},\ \bibinfo {pages} {1153} (\bibinfo {year}
  {1983}{\natexlab{b}})}\BibitemShut {NoStop}%
\bibitem [{\citenamefont {den Nijs}\ and\ \citenamefont
  {Rommelse}(1989)}]{PhysRevB.40.4709}%
  \BibitemOpen
  \bibfield  {author} {\bibinfo {author} {\bibfnamefont {M.}~\bibnamefont {den
  Nijs}}\ and\ \bibinfo {author} {\bibfnamefont {K.}~\bibnamefont {Rommelse}},\
  }\href {\doibase 10.1103/PhysRevB.40.4709} {\bibfield  {journal} {\bibinfo
  {journal} {Phys. Rev. B}\ }\textbf {\bibinfo {volume} {40}},\ \bibinfo
  {pages} {4709} (\bibinfo {year} {1989})}\BibitemShut {NoStop}%
\bibitem [{\citenamefont {Calvanese~Strinati}\ \emph
  {et~al.}(2016)\citenamefont {Calvanese~Strinati}, \citenamefont {Mazza},
  \citenamefont {Endres}, \citenamefont {Rossini},\ and\ \citenamefont
  {Fazio}}]{PhysRevB.94.024302}%
  \BibitemOpen
  \bibfield  {author} {\bibinfo {author} {\bibfnamefont {M.}~\bibnamefont
  {Calvanese~Strinati}}, \bibinfo {author} {\bibfnamefont {L.}~\bibnamefont
  {Mazza}}, \bibinfo {author} {\bibfnamefont {M.}~\bibnamefont {Endres}},
  \bibinfo {author} {\bibfnamefont {D.}~\bibnamefont {Rossini}}, \ and\
  \bibinfo {author} {\bibfnamefont {R.}~\bibnamefont {Fazio}},\ }\href
  {\doibase 10.1103/PhysRevB.94.024302} {\bibfield  {journal} {\bibinfo
  {journal} {Phys. Rev. B}\ }\textbf {\bibinfo {volume} {94}},\ \bibinfo
  {pages} {024302} (\bibinfo {year} {2016})}\BibitemShut {NoStop}%
\bibitem [{\citenamefont {Mazza}\ \emph {et~al.}(2014)\citenamefont {Mazza},
  \citenamefont {Rossini}, \citenamefont {Endres},\ and\ \citenamefont
  {Fazio}}]{PhysRevB.90.020301}%
  \BibitemOpen
  \bibfield  {author} {\bibinfo {author} {\bibfnamefont {L.}~\bibnamefont
  {Mazza}}, \bibinfo {author} {\bibfnamefont {D.}~\bibnamefont {Rossini}},
  \bibinfo {author} {\bibfnamefont {M.}~\bibnamefont {Endres}}, \ and\ \bibinfo
  {author} {\bibfnamefont {R.}~\bibnamefont {Fazio}},\ }\href {\doibase
  10.1103/PhysRevB.90.020301} {\bibfield  {journal} {\bibinfo  {journal} {Phys.
  Rev. B}\ }\textbf {\bibinfo {volume} {90}},\ \bibinfo {pages} {020301}
  (\bibinfo {year} {2014})}\BibitemShut {NoStop}%
\bibitem [{\citenamefont {Strinati}\ \emph {et~al.}(2017)\citenamefont
  {Strinati}, \citenamefont {Rossini}, \citenamefont {Fazio},\ and\
  \citenamefont {Russomanno}}]{PhysRevB.96.214206}%
  \BibitemOpen
  \bibfield  {author} {\bibinfo {author} {\bibfnamefont {M.~C.}\ \bibnamefont
  {Strinati}}, \bibinfo {author} {\bibfnamefont {D.}~\bibnamefont {Rossini}},
  \bibinfo {author} {\bibfnamefont {R.}~\bibnamefont {Fazio}}, \ and\ \bibinfo
  {author} {\bibfnamefont {A.}~\bibnamefont {Russomanno}},\ }\href {\doibase
  10.1103/PhysRevB.96.214206} {\bibfield  {journal} {\bibinfo  {journal} {Phys.
  Rev. B}\ }\textbf {\bibinfo {volume} {96}},\ \bibinfo {pages} {214206}
  (\bibinfo {year} {2017})}\BibitemShut {NoStop}%
\bibitem [{\citenamefont {McGinley}\ and\ \citenamefont
  {Cooper}(2018)}]{PhysRevLett.121.090401}%
  \BibitemOpen
  \bibfield  {author} {\bibinfo {author} {\bibfnamefont {M.}~\bibnamefont
  {McGinley}}\ and\ \bibinfo {author} {\bibfnamefont {N.~R.}\ \bibnamefont
  {Cooper}},\ }\href {\doibase 10.1103/PhysRevLett.121.090401} {\bibfield
  {journal} {\bibinfo  {journal} {Phys. Rev. Lett.}\ }\textbf {\bibinfo
  {volume} {121}},\ \bibinfo {pages} {090401} (\bibinfo {year}
  {2018})}\BibitemShut {NoStop}%
\bibitem [{\citenamefont {Mezio}\ \emph {et~al.}(2015)\citenamefont {Mezio},
  \citenamefont {Lobos}, \citenamefont {Dobry},\ and\ \citenamefont
  {Gazza}}]{PhysRevB.92.205128}%
  \BibitemOpen
  \bibfield  {author} {\bibinfo {author} {\bibfnamefont {A.}~\bibnamefont
  {Mezio}}, \bibinfo {author} {\bibfnamefont {A.~M.}\ \bibnamefont {Lobos}},
  \bibinfo {author} {\bibfnamefont {A.~O.}\ \bibnamefont {Dobry}}, \ and\
  \bibinfo {author} {\bibfnamefont {C.~J.}\ \bibnamefont {Gazza}},\ }\href
  {\doibase 10.1103/PhysRevB.92.205128} {\bibfield  {journal} {\bibinfo
  {journal} {Phys. Rev. B}\ }\textbf {\bibinfo {volume} {92}},\ \bibinfo
  {pages} {205128} (\bibinfo {year} {2015})}\BibitemShut {NoStop}%
\bibitem [{\citenamefont {Alexandrov}\ and\ \citenamefont
  {Coleman}(2014)}]{Coleman2014}%
  \BibitemOpen
  \bibfield  {author} {\bibinfo {author} {\bibfnamefont {V.}~\bibnamefont
  {Alexandrov}}\ and\ \bibinfo {author} {\bibfnamefont {P.}~\bibnamefont
  {Coleman}},\ }\href@noop {} {\bibfield  {journal} {\bibinfo  {journal} {Phys.
  Rev. B}\ }\textbf {\bibinfo {volume} {90}},\ \bibinfo {pages} {115147}
  (\bibinfo {year} {2014})}\BibitemShut {NoStop}%
\bibitem [{\citenamefont {Wolgast}\ \emph {et~al.}(2013)\citenamefont
  {Wolgast}, \citenamefont {Kurdak}, \citenamefont {Sun}, \citenamefont
  {Allen}, \citenamefont {Kim},\ and\ \citenamefont {Fisk}}]{Fiskprb2013}%
  \BibitemOpen
  \bibfield  {author} {\bibinfo {author} {\bibfnamefont {S.}~\bibnamefont
  {Wolgast}}, \bibinfo {author} {\bibfnamefont {C.}~\bibnamefont {Kurdak}},
  \bibinfo {author} {\bibfnamefont {K.}~\bibnamefont {Sun}}, \bibinfo {author}
  {\bibfnamefont {J.~W.}\ \bibnamefont {Allen}}, \bibinfo {author}
  {\bibfnamefont {D.-J.}\ \bibnamefont {Kim}}, \ and\ \bibinfo {author}
  {\bibfnamefont {Z.}~\bibnamefont {Fisk}},\ }\href@noop {} {\bibfield
  {journal} {\bibinfo  {journal} {Phys. Rev. B}\ }\textbf {\bibinfo {volume}
  {88}},\ \bibinfo {pages} {180405} (\bibinfo {year} {2013})}\BibitemShut
  {NoStop}%
\bibitem [{\citenamefont {Zhang}\ \emph {et~al.}(2013)\citenamefont {Zhang},
  \citenamefont {Butch}, \citenamefont {Syers}, \citenamefont {Ziemak},
  \citenamefont {Greene},\ and\ \citenamefont {Paglione}}]{Paglioneprx2013}%
  \BibitemOpen
  \bibfield  {author} {\bibinfo {author} {\bibfnamefont {X.}~\bibnamefont
  {Zhang}}, \bibinfo {author} {\bibfnamefont {N.~P.}\ \bibnamefont {Butch}},
  \bibinfo {author} {\bibfnamefont {P.}~\bibnamefont {Syers}}, \bibinfo
  {author} {\bibfnamefont {S.}~\bibnamefont {Ziemak}}, \bibinfo {author}
  {\bibfnamefont {R.~L.}\ \bibnamefont {Greene}}, \ and\ \bibinfo {author}
  {\bibfnamefont {J.}~\bibnamefont {Paglione}},\ }\href@noop {} {\bibfield
  {journal} {\bibinfo  {journal} {Phys. Rev. X}\ }\textbf {\bibinfo {volume}
  {3}},\ \bibinfo {pages} {011011} (\bibinfo {year} {2013})}\BibitemShut
  {NoStop}%
\bibitem [{\citenamefont {Kim}\ \emph {et~al.}(2013)\citenamefont {Kim},
  \citenamefont {Thomas}, \citenamefont {Grant}, \citenamefont {Botimer},
  \citenamefont {Fisk},\ and\ \citenamefont {Xia}}]{Kim2013}%
  \BibitemOpen
  \bibfield  {author} {\bibinfo {author} {\bibfnamefont {D.~J.}\ \bibnamefont
  {Kim}}, \bibinfo {author} {\bibfnamefont {S.}~\bibnamefont {Thomas}},
  \bibinfo {author} {\bibfnamefont {T.}~\bibnamefont {Grant}}, \bibinfo
  {author} {\bibfnamefont {J.}~\bibnamefont {Botimer}}, \bibinfo {author}
  {\bibfnamefont {Z.}~\bibnamefont {Fisk}}, \ and\ \bibinfo {author}
  {\bibfnamefont {J.}~\bibnamefont {Xia}},\ }\href@noop {} {\bibfield
  {journal} {\bibinfo  {journal} {Sci. Rep.}\ }\textbf {\bibinfo {volume}
  {3}},\ \bibinfo {pages} {3150} (\bibinfo {year} {2013})}\BibitemShut
  {NoStop}%
\bibitem [{\citenamefont {Lobos}\ \emph {et~al.}(2015)\citenamefont {Lobos},
  \citenamefont {Dobry},\ and\ \citenamefont {Galitski}}]{Galitskiprx2015}%
  \BibitemOpen
  \bibfield  {author} {\bibinfo {author} {\bibfnamefont {A.~M.}\ \bibnamefont
  {Lobos}}, \bibinfo {author} {\bibfnamefont {A.~O.}\ \bibnamefont {Dobry}}, \
  and\ \bibinfo {author} {\bibfnamefont {V.}~\bibnamefont {Galitski}},\
  }\href@noop {} {\bibfield  {journal} {\bibinfo  {journal} {Phys. Rev. X}\
  }\textbf {\bibinfo {volume} {5}},\ \bibinfo {pages} {021017} (\bibinfo {year}
  {2015})}\BibitemShut {NoStop}%
\bibitem [{\citenamefont {Hagym\'asi}\ and\ \citenamefont
  {Legeza}(2016)}]{PhysRevB.93.165104}%
  \BibitemOpen
  \bibfield  {author} {\bibinfo {author} {\bibfnamefont {I.}~\bibnamefont
  {Hagym\'asi}}\ and\ \bibinfo {author} {\bibfnamefont {{\"O}.}~\bibnamefont
  {Legeza}},\ }\href {\doibase 10.1103/PhysRevB.93.165104} {\bibfield
  {journal} {\bibinfo  {journal} {Phys. Rev. B}\ }\textbf {\bibinfo {volume}
  {93}},\ \bibinfo {pages} {165104} (\bibinfo {year} {2016})}\BibitemShut
  {NoStop}%
\bibitem [{\citenamefont {Lisandrini}\ \emph {et~al.}(2016)\citenamefont
  {Lisandrini}, \citenamefont {Lobos}, \citenamefont {Dobry},\ and\
  \citenamefont {Gazza}}]{PIP336}%
  \BibitemOpen
  \bibfield  {author} {\bibinfo {author} {\bibfnamefont {F.}~\bibnamefont
  {Lisandrini}}, \bibinfo {author} {\bibfnamefont {A.}~\bibnamefont {Lobos}},
  \bibinfo {author} {\bibfnamefont {A.}~\bibnamefont {Dobry}}, \ and\ \bibinfo
  {author} {\bibfnamefont {C.}~\bibnamefont {Gazza}},\ }\href
  {http://www.papersinphysics.org/papersinphysics/article/view/336} {\bibfield
  {journal} {\bibinfo  {journal} {Pap. Phys.}\ }\textbf {\bibinfo {volume}
  {8}},\ \bibinfo {pages} {080005} (\bibinfo {year} {2016})}\BibitemShut
  {NoStop}%
\bibitem [{\citenamefont {Lisandrini}\ \emph {et~al.}(2017)\citenamefont
  {Lisandrini}, \citenamefont {Lobos}, \citenamefont {Dobry},\ and\
  \citenamefont {Gazza}}]{PhysRevB.96.075124}%
  \BibitemOpen
  \bibfield  {author} {\bibinfo {author} {\bibfnamefont {F.~T.}\ \bibnamefont
  {Lisandrini}}, \bibinfo {author} {\bibfnamefont {A.~M.}\ \bibnamefont
  {Lobos}}, \bibinfo {author} {\bibfnamefont {A.~O.}\ \bibnamefont {Dobry}}, \
  and\ \bibinfo {author} {\bibfnamefont {C.~J.}\ \bibnamefont {Gazza}},\ }\href
  {\doibase 10.1103/PhysRevB.96.075124} {\bibfield  {journal} {\bibinfo
  {journal} {Phys. Rev. B}\ }\textbf {\bibinfo {volume} {96}},\ \bibinfo
  {pages} {075124} (\bibinfo {year} {2017})}\BibitemShut {NoStop}%
\bibitem [{\citenamefont {Pillay}\ and\ \citenamefont
  {McCulloch}(2018)}]{pillay2018topological}%
  \BibitemOpen
  \bibfield  {author} {\bibinfo {author} {\bibfnamefont {J.~C.}\ \bibnamefont
  {Pillay}}\ and\ \bibinfo {author} {\bibfnamefont {I.~P.}\ \bibnamefont
  {McCulloch}},\ }\href {\doibase 10.1103/PhysRevB.97.205133} {\bibfield
  {journal} {\bibinfo  {journal} {Phys. Rev. B}\ }\textbf {\bibinfo {volume}
  {97}},\ \bibinfo {pages} {205133} (\bibinfo {year} {2018})}\BibitemShut
  {NoStop}%
\bibitem [{\citenamefont {Zhong}\ \emph {et~al.}(2017)\citenamefont {Zhong},
  \citenamefont {Liu},\ and\ \citenamefont {Luo}}]{Zhong2017}%
  \BibitemOpen
  \bibfield  {author} {\bibinfo {author} {\bibfnamefont {Y.}~\bibnamefont
  {Zhong}}, \bibinfo {author} {\bibfnamefont {Y.}~\bibnamefont {Liu}}, \ and\
  \bibinfo {author} {\bibfnamefont {H.-G.}\ \bibnamefont {Luo}},\ }\href
  {\doibase 10.1140/epjb/e2017-80102-0} {\bibfield  {journal} {\bibinfo
  {journal} {Eur. Phys. J. B}\ }\textbf {\bibinfo {volume} {90}},\ \bibinfo
  {pages} {147} (\bibinfo {year} {2017})}\BibitemShut {NoStop}%
\bibitem [{\citenamefont {Zhong}\ \emph {et~al.}(2018)\citenamefont {Zhong},
  \citenamefont {Wang}, \citenamefont {Liu}, \citenamefont {Song},
  \citenamefont {Liu},\ and\ \citenamefont {Luo}}]{zhong2017finite}%
  \BibitemOpen
  \bibfield  {author} {\bibinfo {author} {\bibfnamefont {Y.}~\bibnamefont
  {Zhong}}, \bibinfo {author} {\bibfnamefont {Q.}~\bibnamefont {Wang}},
  \bibinfo {author} {\bibfnamefont {Y.}~\bibnamefont {Liu}}, \bibinfo {author}
  {\bibfnamefont {H.-F.}\ \bibnamefont {Song}}, \bibinfo {author}
  {\bibfnamefont {K.}~\bibnamefont {Liu}}, \ and\ \bibinfo {author}
  {\bibfnamefont {H.-G.}\ \bibnamefont {Luo}},\ }\href {\doibase
  10.1007/s11467-018-0868-x} {\bibfield  {journal} {\bibinfo  {journal} {Front.
  Phys.}\ }\textbf {\bibinfo {volume} {14}},\ \bibinfo {pages} {23602}
  (\bibinfo {year} {2018})}\BibitemShut {NoStop}%
\bibitem [{\citenamefont {Bloch}\ \emph {et~al.}(2008)\citenamefont {Bloch},
  \citenamefont {Dalibard},\ and\ \citenamefont {Zwerger}}]{RevModPhys.80.885}%
  \BibitemOpen
  \bibfield  {author} {\bibinfo {author} {\bibfnamefont {I.}~\bibnamefont
  {Bloch}}, \bibinfo {author} {\bibfnamefont {J.}~\bibnamefont {Dalibard}}, \
  and\ \bibinfo {author} {\bibfnamefont {W.}~\bibnamefont {Zwerger}},\ }\href
  {\doibase 10.1103/RevModPhys.80.885} {\bibfield  {journal} {\bibinfo
  {journal} {Rev. Mod. Phys.}\ }\textbf {\bibinfo {volume} {80}},\ \bibinfo
  {pages} {885} (\bibinfo {year} {2008})}\BibitemShut {NoStop}%
\bibitem [{\citenamefont {Haegeman}\ \emph {et~al.}(2011)\citenamefont
  {Haegeman}, \citenamefont {Cirac}, \citenamefont {Osborne}, \citenamefont
  {Pi\ifmmode~\check{z}\else \v{z}\fi{}orn}, \citenamefont {Verschelde},\ and\
  \citenamefont {Verstraete}}]{PhysRevLett.107.070601}%
  \BibitemOpen
  \bibfield  {author} {\bibinfo {author} {\bibfnamefont {J.}~\bibnamefont
  {Haegeman}}, \bibinfo {author} {\bibfnamefont {J.~I.}\ \bibnamefont {Cirac}},
  \bibinfo {author} {\bibfnamefont {T.~J.}\ \bibnamefont {Osborne}}, \bibinfo
  {author} {\bibfnamefont {I.}~\bibnamefont {Pi\ifmmode~\check{z}\else
  \v{z}\fi{}orn}}, \bibinfo {author} {\bibfnamefont {H.}~\bibnamefont
  {Verschelde}}, \ and\ \bibinfo {author} {\bibfnamefont {F.}~\bibnamefont
  {Verstraete}},\ }\href {\doibase 10.1103/PhysRevLett.107.070601} {\bibfield
  {journal} {\bibinfo  {journal} {Phys. Rev. Lett.}\ }\textbf {\bibinfo
  {volume} {107}},\ \bibinfo {pages} {070601} (\bibinfo {year}
  {2011})}\BibitemShut {NoStop}%
\bibitem [{\citenamefont {Haegeman}\ \emph {et~al.}(2016)\citenamefont
  {Haegeman}, \citenamefont {Lubich}, \citenamefont {Oseledets}, \citenamefont
  {Vandereycken},\ and\ \citenamefont {Verstraete}}]{PhysRevB.94.165116}%
  \BibitemOpen
  \bibfield  {author} {\bibinfo {author} {\bibfnamefont {J.}~\bibnamefont
  {Haegeman}}, \bibinfo {author} {\bibfnamefont {C.}~\bibnamefont {Lubich}},
  \bibinfo {author} {\bibfnamefont {I.}~\bibnamefont {Oseledets}}, \bibinfo
  {author} {\bibfnamefont {B.}~\bibnamefont {Vandereycken}}, \ and\ \bibinfo
  {author} {\bibfnamefont {F.}~\bibnamefont {Verstraete}},\ }\href {\doibase
  10.1103/PhysRevB.94.165116} {\bibfield  {journal} {\bibinfo  {journal} {Phys.
  Rev. B}\ }\textbf {\bibinfo {volume} {94}},\ \bibinfo {pages} {165116}
  (\bibinfo {year} {2016})}\BibitemShut {NoStop}%
\bibitem [{\citenamefont {Schollw{\"o}ck}(2011)}]{SCHOLLWOCK201196}%
  \BibitemOpen
  \bibfield  {author} {\bibinfo {author} {\bibfnamefont {U.}~\bibnamefont
  {Schollw{\"o}ck}},\ }\href {\doibase
  https://doi.org/10.1016/j.aop.2010.09.012} {\bibfield  {journal} {\bibinfo
  {journal} {Ann. Phys.}\ }\textbf {\bibinfo {volume} {326}},\ \bibinfo {pages}
  {96 } (\bibinfo {year} {2011})}\BibitemShut {NoStop}%
\bibitem [{\citenamefont {Chiara}\ \emph {et~al.}(2006)\citenamefont {Chiara},
  \citenamefont {Montangero}, \citenamefont {Calabrese},\ and\ \citenamefont
  {Fazio}}]{Fazio2006}%
  \BibitemOpen
  \bibfield  {author} {\bibinfo {author} {\bibfnamefont {G.~D.}\ \bibnamefont
  {Chiara}}, \bibinfo {author} {\bibfnamefont {S.}~\bibnamefont {Montangero}},
  \bibinfo {author} {\bibfnamefont {P.}~\bibnamefont {Calabrese}}, \ and\
  \bibinfo {author} {\bibfnamefont {R.}~\bibnamefont {Fazio}},\ }\href
  {http://stacks.iop.org/1742-5468/2006/i=03/a=P03001} {\bibfield  {journal}
  {\bibinfo  {journal} {J. Stat. Mech.: Theory and Experiment}\ }\textbf
  {\bibinfo {volume} {2006}},\ \bibinfo {pages} {P03001} (\bibinfo {year}
  {2006})}\BibitemShut {NoStop}%
\bibitem [{\citenamefont {White}(1992)}]{White:DMRG1}%
  \BibitemOpen
  \bibfield  {author} {\bibinfo {author} {\bibfnamefont {S.~R.}\ \bibnamefont
  {White}},\ }\href@noop {} {\bibfield  {journal} {\bibinfo  {journal} {Phys.
  Rev. Lett.}\ }\textbf {\bibinfo {volume} {69}},\ \bibinfo {pages} {2863}
  (\bibinfo {year} {1992})}\BibitemShut {NoStop}%
\bibitem [{\citenamefont {White}(1993)}]{White:DMRG2}%
  \BibitemOpen
  \bibfield  {author} {\bibinfo {author} {\bibfnamefont {S.~R.}\ \bibnamefont
  {White}},\ }\href@noop {} {\bibfield  {journal} {\bibinfo  {journal} {Phys.
  Rev. B}\ }\textbf {\bibinfo {volume} {48}},\ \bibinfo {pages} {10345}
  (\bibinfo {year} {1993})}\BibitemShut {NoStop}%
\bibitem [{\citenamefont {Schollw\"ock}(2005)}]{schollwock2005}%
  \BibitemOpen
  \bibfield  {author} {\bibinfo {author} {\bibfnamefont {U.}~\bibnamefont
  {Schollw\"ock}},\ }\href@noop {} {\bibfield  {journal} {\bibinfo  {journal}
  {Rev. Mod. Phys.}\ }\textbf {\bibinfo {volume} {77}},\ \bibinfo {pages} {259}
  (\bibinfo {year} {2005})}\BibitemShut {NoStop}%
\bibitem [{\citenamefont {Hallberg}(2006)}]{hallberg2006}%
  \BibitemOpen
  \bibfield  {author} {\bibinfo {author} {\bibfnamefont {K.}~\bibnamefont
  {Hallberg}},\ }\href@noop {} {\bibfield  {journal} {\bibinfo  {journal} {Adv.
  Phys.}\ }\textbf {\bibinfo {volume} {55}},\ \bibinfo {pages} {477} (\bibinfo
  {year} {2006})}\BibitemShut {NoStop}%
\bibitem [{\citenamefont {Hubig}\ \emph {et~al.}(2015)\citenamefont {Hubig},
  \citenamefont {McCulloch}, \citenamefont {Schollw\"ock},\ and\ \citenamefont
  {Wolf}}]{hubig15:_stric_dmrg}%
  \BibitemOpen
  \bibfield  {author} {\bibinfo {author} {\bibfnamefont {C.}~\bibnamefont
  {Hubig}}, \bibinfo {author} {\bibfnamefont {I.~P.}\ \bibnamefont
  {McCulloch}}, \bibinfo {author} {\bibfnamefont {U.}~\bibnamefont
  {Schollw\"ock}}, \ and\ \bibinfo {author} {\bibfnamefont {F.~A.}\
  \bibnamefont {Wolf}},\ }\href {\doibase 10.1103/PhysRevB.91.155115}
  {\bibfield  {journal} {\bibinfo  {journal} {Phys. Rev. B}\ }\textbf {\bibinfo
  {volume} {91}},\ \bibinfo {pages} {155115} (\bibinfo {year}
  {2015})}\BibitemShut {NoStop}%
\bibitem [{\citenamefont {Verstraete}\ \emph {et~al.}(2004)\citenamefont
  {Verstraete}, \citenamefont {Garc\'{\i}a-Ripoll},\ and\ \citenamefont
  {Cirac}}]{PhysRevLett.93.207204}%
  \BibitemOpen
  \bibfield  {author} {\bibinfo {author} {\bibfnamefont {F.}~\bibnamefont
  {Verstraete}}, \bibinfo {author} {\bibfnamefont {J.~J.}\ \bibnamefont
  {Garc\'{\i}a-Ripoll}}, \ and\ \bibinfo {author} {\bibfnamefont {J.~I.}\
  \bibnamefont {Cirac}},\ }\href {\doibase 10.1103/PhysRevLett.93.207204}
  {\bibfield  {journal} {\bibinfo  {journal} {Phys. Rev. Lett.}\ }\textbf
  {\bibinfo {volume} {93}},\ \bibinfo {pages} {207204} (\bibinfo {year}
  {2004})}\BibitemShut {NoStop}%
\bibitem [{\citenamefont {Ronzheimer}\ \emph {et~al.}(2013)\citenamefont
  {Ronzheimer}, \citenamefont {Schreiber}, \citenamefont {Braun}, \citenamefont
  {Hodgman}, \citenamefont {Langer}, \citenamefont {McCulloch}, \citenamefont
  {Heidrich-Meisner}, \citenamefont {Bloch},\ and\ \citenamefont
  {Schneider}}]{PhysRevLett.110.205301}%
  \BibitemOpen
  \bibfield  {author} {\bibinfo {author} {\bibfnamefont {J.~P.}\ \bibnamefont
  {Ronzheimer}}, \bibinfo {author} {\bibfnamefont {M.}~\bibnamefont
  {Schreiber}}, \bibinfo {author} {\bibfnamefont {S.}~\bibnamefont {Braun}},
  \bibinfo {author} {\bibfnamefont {S.~S.}\ \bibnamefont {Hodgman}}, \bibinfo
  {author} {\bibfnamefont {S.}~\bibnamefont {Langer}}, \bibinfo {author}
  {\bibfnamefont {I.~P.}\ \bibnamefont {McCulloch}}, \bibinfo {author}
  {\bibfnamefont {F.}~\bibnamefont {Heidrich-Meisner}}, \bibinfo {author}
  {\bibfnamefont {I.}~\bibnamefont {Bloch}}, \ and\ \bibinfo {author}
  {\bibfnamefont {U.}~\bibnamefont {Schneider}},\ }\href {\doibase
  10.1103/PhysRevLett.110.205301} {\bibfield  {journal} {\bibinfo  {journal}
  {Phys. Rev. Lett.}\ }\textbf {\bibinfo {volume} {110}},\ \bibinfo {pages}
  {205301} (\bibinfo {year} {2013})}\BibitemShut {NoStop}%
\bibitem [{\citenamefont {Strohmaier}\ \emph {et~al.}(2010)\citenamefont
  {Strohmaier}, \citenamefont {Greif}, \citenamefont {J\"ordens}, \citenamefont
  {Tarruell}, \citenamefont {Moritz}, \citenamefont {Esslinger}, \citenamefont
  {Sensarma}, \citenamefont {Pekker}, \citenamefont {Altman},\ and\
  \citenamefont {Demler}}]{PhysRevLett.104.080401}%
  \BibitemOpen
  \bibfield  {author} {\bibinfo {author} {\bibfnamefont {N.}~\bibnamefont
  {Strohmaier}}, \bibinfo {author} {\bibfnamefont {D.}~\bibnamefont {Greif}},
  \bibinfo {author} {\bibfnamefont {R.}~\bibnamefont {J\"ordens}}, \bibinfo
  {author} {\bibfnamefont {L.}~\bibnamefont {Tarruell}}, \bibinfo {author}
  {\bibfnamefont {H.}~\bibnamefont {Moritz}}, \bibinfo {author} {\bibfnamefont
  {T.}~\bibnamefont {Esslinger}}, \bibinfo {author} {\bibfnamefont
  {R.}~\bibnamefont {Sensarma}}, \bibinfo {author} {\bibfnamefont
  {D.}~\bibnamefont {Pekker}}, \bibinfo {author} {\bibfnamefont
  {E.}~\bibnamefont {Altman}}, \ and\ \bibinfo {author} {\bibfnamefont
  {E.}~\bibnamefont {Demler}},\ }\href {\doibase
  10.1103/PhysRevLett.104.080401} {\bibfield  {journal} {\bibinfo  {journal}
  {Phys. Rev. Lett.}\ }\textbf {\bibinfo {volume} {104}},\ \bibinfo {pages}
  {080401} (\bibinfo {year} {2010})}\BibitemShut {NoStop}%
\bibitem [{\citenamefont {Pollmann}\ \emph {et~al.}(2010)\citenamefont
  {Pollmann}, \citenamefont {Turner}, \citenamefont {Berg},\ and\ \citenamefont
  {Oshikawa}}]{Pollmann2010prb}%
  \BibitemOpen
  \bibfield  {author} {\bibinfo {author} {\bibfnamefont {F.}~\bibnamefont
  {Pollmann}}, \bibinfo {author} {\bibfnamefont {A.~M.}\ \bibnamefont
  {Turner}}, \bibinfo {author} {\bibfnamefont {E.}~\bibnamefont {Berg}}, \ and\
  \bibinfo {author} {\bibfnamefont {M.}~\bibnamefont {Oshikawa}},\ }\href@noop
  {} {\bibfield  {journal} {\bibinfo  {journal} {Phys. Rev. B}\ }\textbf
  {\bibinfo {volume} {81}},\ \bibinfo {pages} {064439} (\bibinfo {year}
  {2010})}\BibitemShut {NoStop}%
\bibitem [{\citenamefont {Turner}\ \emph {et~al.}(2011)\citenamefont {Turner},
  \citenamefont {Pollmann},\ and\ \citenamefont {Berg}}]{Pollmann2011prb}%
  \BibitemOpen
  \bibfield  {author} {\bibinfo {author} {\bibfnamefont {A.~M.}\ \bibnamefont
  {Turner}}, \bibinfo {author} {\bibfnamefont {F.}~\bibnamefont {Pollmann}}, \
  and\ \bibinfo {author} {\bibfnamefont {E.}~\bibnamefont {Berg}},\ }\href@noop
  {} {\bibfield  {journal} {\bibinfo  {journal} {Phys. Rev. B}\ }\textbf
  {\bibinfo {volume} {83}},\ \bibinfo {pages} {075102} (\bibinfo {year}
  {2011})}\BibitemShut {NoStop}%
\bibitem [{\citenamefont {Pollmann}\ and\ \citenamefont
  {Turner}(2012)}]{Pollmann2012prb}%
  \BibitemOpen
  \bibfield  {author} {\bibinfo {author} {\bibfnamefont {F.}~\bibnamefont
  {Pollmann}}\ and\ \bibinfo {author} {\bibfnamefont {A.~M.}\ \bibnamefont
  {Turner}},\ }\href@noop {} {\bibfield  {journal} {\bibinfo  {journal} {Phys.
  Rev. B}\ }\textbf {\bibinfo {volume} {86}},\ \bibinfo {pages} {125441}
  (\bibinfo {year} {2012})}\BibitemShut {NoStop}%
\bibitem [{\citenamefont {Cheuk}\ \emph {et~al.}(2016)\citenamefont {Cheuk},
  \citenamefont {Nichols}, \citenamefont {Lawrence}, \citenamefont {Okan},
  \citenamefont {Zhang}, \citenamefont {Khatami}, \citenamefont {Trivedi},
  \citenamefont {Paiva}, \citenamefont {Rigol},\ and\ \citenamefont
  {Zwierlein}}]{Cheuk1260}%
  \BibitemOpen
  \bibfield  {author} {\bibinfo {author} {\bibfnamefont {L.~W.}\ \bibnamefont
  {Cheuk}}, \bibinfo {author} {\bibfnamefont {M.~A.}\ \bibnamefont {Nichols}},
  \bibinfo {author} {\bibfnamefont {K.~R.}\ \bibnamefont {Lawrence}}, \bibinfo
  {author} {\bibfnamefont {M.}~\bibnamefont {Okan}}, \bibinfo {author}
  {\bibfnamefont {H.}~\bibnamefont {Zhang}}, \bibinfo {author} {\bibfnamefont
  {E.}~\bibnamefont {Khatami}}, \bibinfo {author} {\bibfnamefont
  {N.}~\bibnamefont {Trivedi}}, \bibinfo {author} {\bibfnamefont
  {T.}~\bibnamefont {Paiva}}, \bibinfo {author} {\bibfnamefont
  {M.}~\bibnamefont {Rigol}}, \ and\ \bibinfo {author} {\bibfnamefont {M.~W.}\
  \bibnamefont {Zwierlein}},\ }\href {\doibase 10.1126/science.aag3349}
  {\bibfield  {journal} {\bibinfo  {journal} {Science}\ }\textbf {\bibinfo
  {volume} {353}},\ \bibinfo {pages} {1260} (\bibinfo {year}
  {2016})}\BibitemShut {NoStop}%
\bibitem [{\citenamefont {Parsons}\ \emph {et~al.}(2016)\citenamefont
  {Parsons}, \citenamefont {Mazurenko}, \citenamefont {Chiu}, \citenamefont
  {Ji}, \citenamefont {Greif},\ and\ \citenamefont {Greiner}}]{Parsons1253}%
  \BibitemOpen
  \bibfield  {author} {\bibinfo {author} {\bibfnamefont {M.~F.}\ \bibnamefont
  {Parsons}}, \bibinfo {author} {\bibfnamefont {A.}~\bibnamefont {Mazurenko}},
  \bibinfo {author} {\bibfnamefont {C.~S.}\ \bibnamefont {Chiu}}, \bibinfo
  {author} {\bibfnamefont {G.}~\bibnamefont {Ji}}, \bibinfo {author}
  {\bibfnamefont {D.}~\bibnamefont {Greif}}, \ and\ \bibinfo {author}
  {\bibfnamefont {M.}~\bibnamefont {Greiner}},\ }\href {\doibase
  10.1126/science.aag1430} {\bibfield  {journal} {\bibinfo  {journal}
  {Science}\ }\textbf {\bibinfo {volume} {353}},\ \bibinfo {pages} {1253}
  (\bibinfo {year} {2016})}\BibitemShut {NoStop}%
\bibitem [{\citenamefont {Cocchi}\ \emph {et~al.}(2016)\citenamefont {Cocchi},
  \citenamefont {Miller}, \citenamefont {Drewes}, \citenamefont {Koschorreck},
  \citenamefont {Pertot}, \citenamefont {Brennecke},\ and\ \citenamefont
  {K\"ohl}}]{PhysRevLett.116.175301}%
  \BibitemOpen
  \bibfield  {author} {\bibinfo {author} {\bibfnamefont {E.}~\bibnamefont
  {Cocchi}}, \bibinfo {author} {\bibfnamefont {L.~A.}\ \bibnamefont {Miller}},
  \bibinfo {author} {\bibfnamefont {J.~H.}\ \bibnamefont {Drewes}}, \bibinfo
  {author} {\bibfnamefont {M.}~\bibnamefont {Koschorreck}}, \bibinfo {author}
  {\bibfnamefont {D.}~\bibnamefont {Pertot}}, \bibinfo {author} {\bibfnamefont
  {F.}~\bibnamefont {Brennecke}}, \ and\ \bibinfo {author} {\bibfnamefont
  {M.}~\bibnamefont {K\"ohl}},\ }\href {\doibase
  10.1103/PhysRevLett.116.175301} {\bibfield  {journal} {\bibinfo  {journal}
  {Phys. Rev. Lett.}\ }\textbf {\bibinfo {volume} {116}},\ \bibinfo {pages}
  {175301} (\bibinfo {year} {2016})}\BibitemShut {NoStop}%
\bibitem [{\citenamefont {Hilker}\ \emph {et~al.}(2017)\citenamefont {Hilker},
  \citenamefont {Salomon}, \citenamefont {Grusdt}, \citenamefont {Omran},
  \citenamefont {Boll}, \citenamefont {Demler}, \citenamefont {Bloch},\ and\
  \citenamefont {Gross}}]{Hilker484}%
  \BibitemOpen
  \bibfield  {author} {\bibinfo {author} {\bibfnamefont {T.~A.}\ \bibnamefont
  {Hilker}}, \bibinfo {author} {\bibfnamefont {G.}~\bibnamefont {Salomon}},
  \bibinfo {author} {\bibfnamefont {F.}~\bibnamefont {Grusdt}}, \bibinfo
  {author} {\bibfnamefont {A.}~\bibnamefont {Omran}}, \bibinfo {author}
  {\bibfnamefont {M.}~\bibnamefont {Boll}}, \bibinfo {author} {\bibfnamefont
  {E.}~\bibnamefont {Demler}}, \bibinfo {author} {\bibfnamefont
  {I.}~\bibnamefont {Bloch}}, \ and\ \bibinfo {author} {\bibfnamefont
  {C.}~\bibnamefont {Gross}},\ }\href {\doibase 10.1126/science.aam8990}
  {\bibfield  {journal} {\bibinfo  {journal} {Science}\ }\textbf {\bibinfo
  {volume} {357}},\ \bibinfo {pages} {484} (\bibinfo {year}
  {2017})}\BibitemShut {NoStop}%
\end{thebibliography}%

\end{document}